\documentclass[reprint,amsmath,amssymb,aps,prl,superscriptaddress,nofootinbib,floatfix]{revtex4-2}

\usepackage{graphicx}
\usepackage{bm}
\usepackage{mathrsfs}
\usepackage{amsmath}
\usepackage{latexsym}
\usepackage{array}
\usepackage{multirow}
\usepackage{booktabs}
\usepackage{tikz}
\usetikzlibrary{patterns,decorations.pathmorphing,shapes}
\usepackage[colorlinks=true, linkcolor=blue, urlcolor=blue, citecolor=magenta]{hyperref}
\graphicspath{{./}{figures/}{../figures/}}

\begin{document}

\title{Depletion Capture and Rayleigh--Onsager Dynamics in Receptor-Mediated Endocytosis}

\author{Jinjie Liu}
\affiliation{Institute of Theoretical Physics, Chinese Academy of Sciences, Beijing 100190, China}
\author{Zhongcan Ouyang}
\email{oy@itp.ac.cn}
\affiliation{Institute of Theoretical Physics, Chinese Academy of Sciences, Beijing 100190, China}
\author{Hao Wu}
\email{wuhao@ucas.ac.cn}
\affiliation{Zhejiang Key Laboratory of Soft Matter Biomedical Materials, Wenzhou Institute,
University of Chinese Academy of Sciences, Wenzhou, Zhejiang 325000, China}

\date{\today}

\begin{abstract}
Receptor-mediated endocytosis requires a particle to approach a membrane in a crowded extracellular environment, form ligand-receptor bonds, and wrap the membrane against bending, tension, and cytoskeletal viscoelastic resistance. We develop a two-stage theory. The capture stage uses the exact ideal hard-sphere Asakura--Oosawa sphere-plane excluded-volume geometry. The wrapping stage combines a spherical-cap coordinate, a quadratic single-mode free energy, a standard linear solid with one internal strain variable, and a two-variable Rayleigh--Onsager (RO) gradient flow. The creep compliance is mapped exactly onto a long-time spring in parallel with a Maxwell branch, preserving both the creep retardation and stress relaxation times. The free energy is strictly convex by its Hessian convexity, and the linear wrapping system has two real negative eigenvalues and an explicit double-exponential solution. Two independent arguments establish strict monotonicity for positive initial drive: the irreducible Metzler structure with its positive semigroup and the interlacing of Laplace poles with positive velocity residues. The first-passage equation has a unique finite root below the unconstrained long-time limit. Normal-cone conditions specify the projected dynamics on $0\le x\le2$, and the projected system retains the Lyapunov inequality of the unconstrained gradient flow. From the same free energy we derive long-time adhesion thresholds, the threshold-minimizing radius, the finite-size window, and the critical cell modulus. The $95\%$ spherical-area criterion is treated as an operational near-complete threshold, with $90\%$, $95\%$, and $99\%$ compared. The capture analysis includes the exact overlap volume, depletion force, Carnahan--Starling pressure rescaling, Stokes approach time, and Brenner wall-drag correction. Our framework yields explicit long-time thresholds, a finite-size window, and a critical cell modulus, and it provides a mathematically rigorous RO description of depletion-assisted capture and viscoelastic wrapping for receptor-mediated endocytosis.
\end{abstract}

\maketitle

\section{Introduction}
\label{sec:intro}

Receptor-mediated endocytosis underlies viral entry, nutrient uptake, and nanoparticle internalization \cite{doherty2009,zhang2015physical,wu2026unveiling}. A particle must first approach the crowded extracellular membrane, form a specific ligand-receptor bond, and then be wrapped while the cytoskeleton deforms and dissipates energy. Continuum models usually start from established contact or a prescribed wrapping coordinate, and receptor diffusion is often treated as rate limiting \cite{giao2005,Shi2008,mercer2010virus,Marsh2006}. However, ligand-receptor binding requires nanometric separation, and the effect of extracellular crowders on the particle distribution near the membrane is unresolved when contact is assumed at the outset.

Recent studies indicate that endocytosis can require an initial affinity \cite{Verma2010}, which may arise from entropic depletion forces generated by smaller molecules \cite{Nelson2004,asakura1954,asakura1958,Vrij1976}. Depletion attraction is well established in colloid science \cite{Dinsmore1995,Imhof1995,Steiner1995,Ilett1995}, has been discussed for vesicles \cite{Dinsmore1998} and mammalian cells \cite{Irajizad2017}, and is reviewed at the nano-bio interface \cite{Wan2024}. Crowding can drive macromolecular association without direct attraction \cite{Batra2009,Gupta2017,Dey2022,ellis2001crowding}, and the entropic effects of polymers anchored on membrane surface can modulate membrane mechanics and domain separation \cite{wu2013mechanical,wu2013effects}. Depletion forces are weaker than specific ligand-receptor binding; their primary role may be to lower the initial contact barrier rather than drive wrapping alone.

\begin{figure}[t]
\centering
\includegraphics[width=\linewidth]{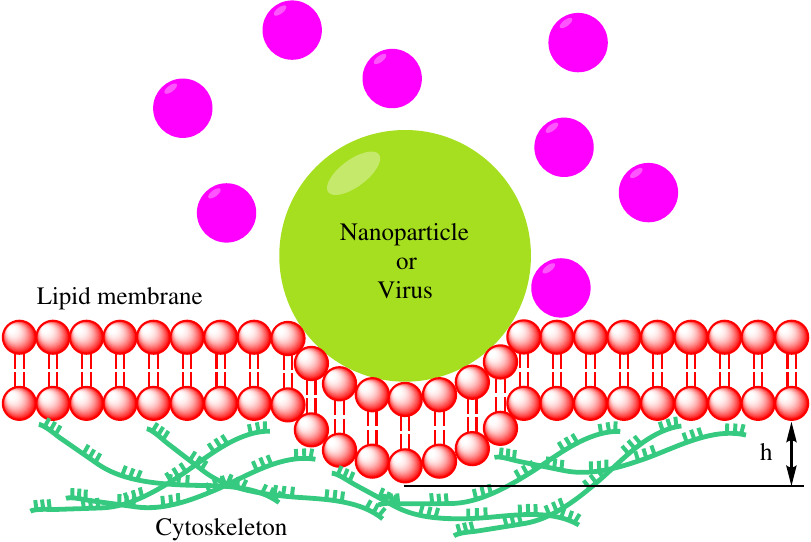}
\caption{Schematic of depletion-assisted receptor-mediated endocytosis. Extracellular crowders (pink) generate an AO depletion attraction that draws the virus/nanoparticle (green) toward the membrane; $h$ is the spherical-cap height from the pole to the contact membrane surface, and the underlying cytoskeleton mediates the viscoelastic wrapping response.}
\label{fig:figure01}
\end{figure}

A key geometric issue is the depletion volume for finite spheres near a planar membrane. The conventional spherical-cap approximation must be expressed in terms of the excluded volume of small-particle centers. Before contact, the exact ideal hard-sphere Asakura--Oosawa (AO) volume follows from the overlap of the particle exclusion sphere of radius $R+r$ and the membrane exclusion half-space $z<r$. Its analytic continuation into the wrapping phase is a separate approximation. We therefore separate the two reductions: the capture stage uses the exact sphere-plane excluded-volume construction, whereas the wrapping stage uses a local parallel-plate adhesion density. The contact energy is not divided by an arbitrary area to define an adhesion density. The two-stage geometry is illustrated in Fig.~\ref{fig:figure01}.

The cell is not purely elastic. The cytoskeleton in a viscous cytosol confers creep and stress relaxation on seconds-to-minutes time scales \cite{mofrad2009cytoskeletal,Rotsch1999}, overlapping some endocytic events \cite{Ryan1996timing,Liou1997autophagic,Elkin2016endocytic}. Elastic half-space or purely elastic descriptions \cite{Sun2006,giao2005} can therefore misrepresent the energy landscape, and viscoelastic resistance can become a kinetic bottleneck. Actin disruption alters viral entry rates \cite{matarrese2005human}; the observed uptake optimum near $50\,\mathrm{nm}$ in diameter (radius about $25\,\mathrm{nm}$) \cite{Chithrani2006} has motivated kinetic and non-equilibrium models. Forces during ventral uptake \cite{WiegandEtAl2020} and viscoelastic properties of basal membranes and cortices \cite{Janshoff2021} have been measured. Previous engulfment models include cytoskeleton deformation after contact \cite{Zhang2021}, and HIV entry modeling similarly assumes prior contact \cite{Kruse2023mathematical}. These studies focus on post-contact wrapping and leave the initial approach and the internal-variable memory of the standard linear solid outside their scope.

The Onsager variational principle couples free-energy storage and viscous dissipation \cite{Onsager1931a,Onsager1931b,Doi2013} and has been applied to polymers, colloids, and membranes \cite{Doi2013,Arroyo2009}. Reduced engulfment descriptions often use a single coordinate $h(t)$ and $\zeta(h)\dot h=F(h)$, while the elastic-viscoelastic correspondence principle extends Hertzian contact to viscoelastic media \cite{Lee1960,Radok1957}. Recently we have developed an Onsager framework for cytoskeletal viscoelastic wrapping \cite{Liu2026e2}, yielding a conditional minimum ligand density, a finite size window, and a stiffness-dependent upper limit. That framework exposes the need for a more complete treatment of viscoelastic memory and first-passage time, which we address here. The creep compliance of the earlier framework is mapped exactly onto a long-time spring in parallel with a Maxwell branch, preserving both the creep retardation time and the stress relaxation time. The resulting two-variable RO gradient flow has an explicit spectrum, a double-exponential solution, strict monotonicity of the wrapping coordinate, a unique first-passage root, and a normal-cone projection on $0\le x\le2$. The earlier scalar description is retained only as a limiting comparison.

The main elements are as follows. (i) The capture stage uses the exact ideal hard-sphere AO sphere-plane overlap volume, not a fitted cap or far-field approximation. (ii) The Carnahan--Starling (CS) replacement is introduced as a pressure rescaling of the ideal overlap geometry; its limitation near hard-sphere freezing is stated, and the high-density endpoint is marked as an extrapolation of the fluid branch. (iii) The wrapping stage uses a spherical-cap coordinate, Helfrich bending \cite{helfrich1973,Canham1970,Ouyang1989,wu2018getting,wu2023,wu2025generalized}, membrane tension, local adhesion, and a single-mode standard linear solid with one internal strain variable. (iv) The creep compliance of the earlier framework is mapped exactly to the internal-variable model, giving $E_0=E_v$, $E_\infty=E_vE_c/(E_v+E_c)$, $E_1=E_v^2/(E_v+E_c)$, $\eta_M=\eta_cE_v^2/(E_v+E_c)^2$, and $\tau_{\mathrm{rel}}=\eta_c/(E_v+E_c)$; the creep retardation time $\tau_c=\eta_c/E_c$ and $\tau_{\mathrm{rel}}$ are distinct. (v) The two-variable linear system has two real negative eigenvalues and an explicit double-exponential solution. Strict monotonicity for positive initial drive follows independently from the irreducible Metzler structure and positive semigroup, and from the interlacing of Laplace poles and positive residues of the velocity. (vi) The first-passage equation has a unique finite root when the target lies below the unconstrained long-time limit; the projected dynamics on $0\le x\le2$ is specified by normal-cone conditions. (vii) Long-time adhesion thresholds, the threshold-minimizing radius, the finite size window, and the critical cell modulus follow from the same free energy. The static threshold-minimizing radius, the time-optimal radius, and the stiffness-peak radius are distinct. (viii) The $95\%$ spherical-area criterion is treated as an operational near-complete threshold, and $90\%$, $95\%$, and $99\%$ results are compared. The parameters used throughout are collected in Table~\ref{tab:params}, and the geometrical and algebraic derivations are collected in Appendices. 

The deterministic approach time considered here is conditional on entry into the depletion layer. The stochastic approach that leads to first-bond formation, including the reaction-before-escape probability and the conditional capture time, is treated separately in the Smoluchowski--Robin (SR) theory of Ref.~\cite{wu2026single}. The subsequent multivalent bond stabilization after the first bond is treated in Ref.~\cite{wu2026multivalent}. The present two-stage description is therefore not coupled through a single reaction coordinate, and it does not attempt to provide a full stochastic total endocytosis time. These distinctions define the scope of the present deterministic-approach results and their interface with the stochastic capture and multivalent-stabilization stages.

\begin{table*}[t]
\caption{Parameters used in the model. References are provided for each parameter.}
\label{tab:params}
\begin{tabular*}{\textwidth}{@{\extracolsep{\fill}}l l l@{}}
\toprule
\textbf{Parameter} & \textbf{Value} & \textbf{References} \\
\midrule
Virus radius \(R\) & \(50\)--\(100\) nm & \cite{Chithrani2006,Burrell2016,WiegandEtAl2020,zhang2015physical} \\
Crowder radius \(r\) & \(5\)--\(20\) nm & \cite{Batra2009,Gupta2017,Kuznetsova2014,Dey2022,ellis2001crowding,Alfano2024,Mitchison2019} \\
Crowder concentration \(c\) & \(1.5\times10^{22}\) m\(^{-3}\) & \cite{Mitchison2019,Alfano2024,Kuznetsova2014} \\
Receptor-ligand complex size \(\delta\) & \(5\) nm & \cite{giao2005,hu2013binding,chen2022quantification} \\
Entry offset \(\epsilon\) & \(0.01\) nm & Chosen for deterministic approach estimate; \(0<\epsilon<2r-\delta\) \\
Total adhesion energy density \(w_{\mathrm{total}}\) & \(2.9\)--\(5.8\times10^{-4}\) J\,m\(^{-2}\) & \cite{Sun2006,chen2022quantification,Yuan2010,Li2012} \\
Boltzmann constant \(k_B\) & \(1.38\times10^{-23}\) J\,K\(^{-1}\) & Fundamental constant \\
Model temperature \(T\) & \(300\) K & Model temperature \\
Membrane bending modulus \(\kappa\) & \(8.18\times10^{-20}\) J & \cite{Li2012,Sun2006} \\
Membrane surface tension \(\gamma\) & \(2.07\times10^{-5}\) N\,m\(^{-1}\) & \cite{Sun2006,Li2024,Dai1998} \\
Water viscosity \(\mu\) & \(10^{-3}\) N\,s\,m\(^{-2}\) & Standard value \\
Cytoskeleton viscosity \(\eta_c\) & \(2696\) N\,s\,m\(^{-2}\) & \cite{Micoulet2005,Rotsch1999,Janshoff2021} \\
Virus Young's modulus \(E_v\) & \(10^8\) N\,m\(^{-2}\) & \cite{Sun2006,Li2012} \\
Cell Young's modulus \(E_c\) & \(2\times10^3\) N\,m\(^{-2}\) & \cite{Micoulet2005,Muenker2024,Rotsch1999,Janshoff2021} \\
Creep retardation time \(\tau=\eta_c/E_c\) & \(1.348\) s & Derived from \(\eta_c/E_c\) \\
\bottomrule
\end{tabular*}
\end{table*}

Section~\hyperref[sec:capture]{II} develops the exact AO sphere-plane geometry, depletion potential and force, CS pressure rescaling, and near-wall approach time with bulk Stokes and Brenner normal drag. Section~\hyperref[sec:wrapping-free-energy]{III} introduces the spherical-cap coordinate and free-energy terms for Helfrich bending, tension, adhesion, and the single-mode standard linear solid. Section~\hyperref[sec:sls-mapping]{IV} maps the creep compliance of \cite{Liu2026e2} to the internal-variable model. Section~\hyperref[sec:gradient-flow]{V} formulates the two-variable RO gradient flow, its spectrum, exact double-exponential solution, and two monotonicity arguments. Section~\hyperref[sec:projected]{VI} treats the projected dynamics on $0\le x\le2$ and the first-passage equation. Section~\hyperref[sec:thresholds]{VII} derives long-time thresholds, the threshold-minimizing radius, the critical cell modulus, the size window, and phase boundaries. Section~\hyperref[sec:criterion95]{VIII} discusses the $95\%$ criterion and its sensitivity. Section~\hyperref[sec:numerics]{IX} reports representative numerical values, and Section~\hyperref[sec:discussion]{X} discusses the model assumptions, their scope, and natural extensions. Appendices collect the geometrical and algebraic derivations.

\section{Depletion capture before contact}
\label{sec:capture}

\subsection{Exact AO sphere-plane excluded volume}
\label{subsec:AO-geometry}

Let the membrane be the plane $z=0$ with the crowder-accessible side $z>0$. A spherical particle of radius $R$ has its lowest point at height $d\ge0$ above the membrane. The particle center is at $z_C=R+d$ \cite{Liu2026e2}. A crowder of radius $r$ cannot approach the particle surface closer than $r$. The particle-induced exclusion region for crowder centers is therefore a sphere of radius
\begin{equation}
A=R+r
\label{eq:Aradius}
\end{equation}
centered at $z_C=R+d$. The wall-induced exclusion region for crowder centers is the half-space $z<r$. For $d<2r$, the intersection of the exclusion sphere and the wall-excluded half-space is a spherical cap with height
\begin{equation}
q=r-(z_C-A)=2r-d.
\label{eq:capheight}
\end{equation}
For $d\ge2r$ the overlap volume is zero. The complete derivation of the overlap volume is given in Appendix~\hyperref[app:AO]{A}.

\begin{figure*}[t]
\centering
\includegraphics[width=\linewidth]{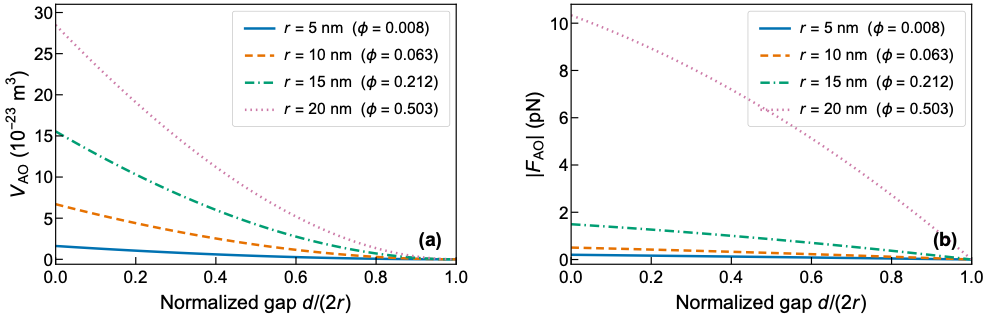}
\caption{Exact AO sphere-plane geometry at $R=50\,\mathrm{nm}$ and Table-I number density $c=1.5\times10^{22}\,\mathrm{m^{-3}}$, using the CS pressure rescaling. (a) Overlap volume $V_{\mathrm{AO}}$ versus normalized gap $d/(2r)$ for $r=5$, $10$, $15$, and $20\,\mathrm{nm}$; each curve decreases monotonically and vanishes quadratically at $d=2r$. (b) Attractive-force magnitude $|F_{\mathrm{AO}}|$ versus $d/(2r)$ for the same radii; the force vanishes linearly at $d=2r$ and reaches its maximum at contact.}
\label{fig:figure02}
\end{figure*}

The cap volume of a sphere of radius $A$ and height $q$ is $\pi q^2(A-q/3)$. Substituting $A=R+r$ and $q=2r-d$ gives the exact ideal hard-sphere AO sphere-plane overlap volume \cite{Liu2026e2}
\begin{equation}
V_{\mathrm{AO}}(d)=\frac{\pi}{3}(2r-d)^2(3R+r+d),\quad 0\le d\le2r,
\label{eq:VAO}
\end{equation}
and $V_{\mathrm{AO}}(d)=0$ for $d\ge2r$. At contact,
\begin{equation}
V_{\mathrm{AO}}(0)
=\frac{4\pi}{3}r^2(3R+r)
=4\pi Rr^2+\frac{4\pi}{3}r^3.
\label{eq:VAOcontact}
\end{equation}
Differentiating \eqref{eq:VAO} gives
\begin{equation}
V_{\mathrm{AO}}'(d)
=-\pi(2r-d)(2R+d)\le0,
\label{eq:dVAO}
\end{equation}
and
\begin{equation}
V_{\mathrm{AO}}''(d)
=2\pi(R-r+d)>0
\label{eq:Vsecond}
\end{equation}
for $R>r$, which holds in the Table-I range. The overlap volume is monotonically decreasing and convex in the physical gap. Near the outer edge $d=2r$, writing $s=2r-d$ gives
\begin{equation}
V_{\mathrm{AO}}(2r-s)
=\pi(R+r)s^2-\frac{\pi}{3}s^3.
\label{eq:VAOedge}
\end{equation}
The volume vanishes quadratically at $s=0$, and the force vanishes linearly. The overlap volume and the attractive-force magnitude are shown in Fig.~\ref{fig:figure02}.

\subsection{Depletion potential and force}
\label{subsec:AO-potential}

For an ideal crowder reservoir with osmotic pressure $\Pi$, the AO free energy is
\begin{equation}
U_{\mathrm{AO}}(d)=-\Pi V_{\mathrm{AO}}(d),\qquad U_{\mathrm{AO}}(d)=0\quad(d\ge2r).
\label{eq:UAO}
\end{equation}
The generalized force along increasing gap is
\begin{equation}
F_d(d)
=-\frac{\mathrm{d}U_{\mathrm{AO}}}{\mathrm{d}d}
=-\pi\Pi(2r-d)(2R+d)\le0.
\label{eq:Fd}
\end{equation}
The negative sign means that the force pushes the particle toward smaller $d$, that is, toward the membrane. The magnitude of the attractive force is
\begin{equation}
|F_{\mathrm{AO}}(d)|=\pi\Pi(2r-d)(2R+d),
\label{eq:FAOmag}
\end{equation}
and at contact
\begin{equation}
|F_{\mathrm{AO}}(0)|=4\pi\Pi Rr.
\label{eq:FAOcontact}
\end{equation}
At the outer edge $d=2r$, both the potential and the force vanish continuously. The depletion layers of the particle and the wall begin to overlap as the particle approaches the membrane; the overlap removes excluded volume from the crowder bath and produces an effective attraction toward the membrane. The force magnitude for several crowder radii is displayed in Fig.~\ref{fig:figure02}(b).

\subsection{CS pressure rescaling and state limitations}
\label{subsec:CS}

The ideal van't Hoff pressure is $\Pi_{\mathrm{id}}=ck_{\mathrm{B}}T$. For a monodisperse hard-sphere fluid, the CS compressibility factor is \cite{carnahan1969}
\begin{equation}
Z_{\mathrm{CS}}(\phi)=\frac{1+\phi+\phi^2-\phi^3}{(1-\phi)^3},
\quad
\phi=\frac{4\pi}{3}cr^3,
\label{eq:ZCS}
\end{equation}
and the CS pressure is
\begin{equation}
\Pi_{\mathrm{CS}}=ck_{\mathrm{B}}TZ_{\mathrm{CS}}(\phi).
\label{eq:PiCS}
\end{equation}
Replacing $\Pi_{\mathrm{id}}$ by $\Pi_{\mathrm{CS}}$ while retaining the ideal AO geometry gives the pressure-rescaled approximation
\begin{equation}
U_{\mathrm{AO}}^{\mathrm{CS}}(d)=-\Pi_{\mathrm{CS}}V_{\mathrm{AO}}(d).
\label{eq:UCS}
\end{equation}
This replacement changes the bulk osmotic pressure and retains the ideal excluded-volume geometry. It does not reproduce finite-density structural correlations, layering, bridging, or curvature corrections of a nonideal depletion potential.

\begin{figure}[t]
\centering
\includegraphics[width=\linewidth]{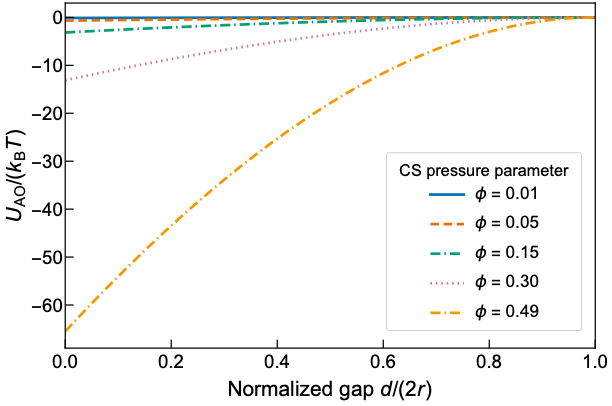}
\caption{Crowding dependence of the AO potential at fixed $R=50\,\mathrm{nm}$, $r=15\,\mathrm{nm}$, and $T=300\,\mathrm{K}$, with the CS pressure rescaling. The volume fraction $\phi=0.01$, $0.05$, $0.15$, $0.30$, and $0.49$ is used as a curve parameter. All curves share the same AO gap shape; the well depth increases strongly as $\phi$ approaches the hard-sphere freezing regime.}
\label{fig:figure03}
\end{figure}

For the representative values $R=50\,\mathrm{nm}$, $r=15\,\mathrm{nm}$, $c=1.5\times10^{22}\,\mathrm{m^{-3}}$, and $T=300\,\mathrm{K}$,
\begin{equation}
\phi=0.2120575,\quad
Z_{\mathrm{CS}}(\phi)=2.5500782,
\label{eq:phirep}
\end{equation}
\begin{equation}
\Pi_{\mathrm{id}}=62.1292\,\mathrm{Pa},
\quad
\Pi_{\mathrm{CS}}=158.434\,\mathrm{Pa}.
\label{eq:Pirep}
\end{equation}
The contact depth is
\begin{equation}
\beta|U_{\mathrm{AO}}(0)|=
\begin{cases}
2.3326, & \text{ideal pressure},\\
5.9484, & \text{CS pressure rescaling},
\end{cases}
\label{eq:contactdepth}
\end{equation}
and the contact volume is
\begin{equation}
V_{\mathrm{AO}}(0)=1.5550884\times10^{-22}\,\mathrm{m^3}.
\label{eq:Vcontactrep}
\end{equation}
The crowding dependence of the AO potential is shown in Fig.~\ref{fig:figure03}. At fixed number density $c$, increasing $r$ changes the volume fraction as $\phi\propto r^3$. At $r=20\,\mathrm{nm}$, $\phi=0.50265$, above the hard-sphere fluid freezing volume fraction $\phi_f\simeq0.494$ \cite{pusey1986,martelozzo2002}. The CS expression remains algebraically defined for $\phi<1$, but for $\phi>\phi_f$ it is used here as an extrapolation of the fluid branch and not as a prediction of a stable uniform monodisperse hard-sphere fluid.

\begin{figure*}[t]
\centering
\includegraphics[width=\linewidth]{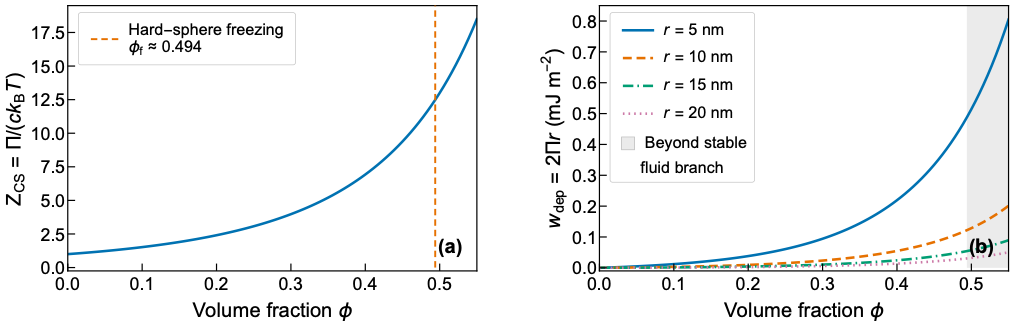}
\caption{CS crowding and planar AO adhesion. (a) Compressibility factor $Z_{\mathrm{CS}}(\phi)$ versus volume fraction; the dashed line marks the hard-sphere fluid freezing fraction $\phi_f\simeq0.494$, and the shaded range $\phi\gtrsim0.494$ is outside the stable monodisperse fluid branch. (b) Planar depletion adhesion density $w_{\mathrm{dep}}=2\Pi r$ versus $\phi$ for $r=5$, $10$, $15$, and $20\,\mathrm{nm}$; at fixed $\phi$, smaller crowders give larger $w_{\mathrm{dep}}$, and all curves rise steeply near freezing.}
\label{fig:figure04}
\end{figure*}

\begin{figure*}[t]
\centering
\includegraphics[width=\linewidth]{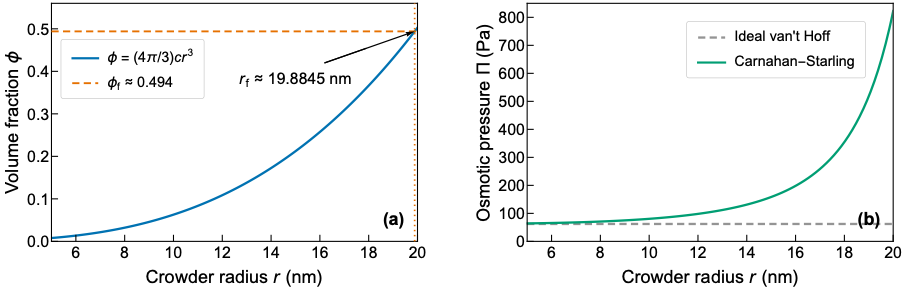}
\caption{Mapping of the Table-I crowder radius $r$ at fixed $c=1.5\times10^{22}\,\mathrm{m^{-3}}$ and $T=300\,\mathrm{K}$. (a) Volume fraction $\phi=(4\pi/3)cr^3$ versus $r$; the dashed line is $\phi_f\simeq0.494$, and the freezing radius is $r_f\approx19.88\,\mathrm{nm}$. (b) Osmotic pressure versus $r$: the ideal van't Hoff pressure is independent of $r$ (dashed horizontal line), whereas the CS pressure increases steeply through $\phi(r)$ and reaches the high-density extrapolation near $r=20\,\mathrm{nm}$.}
\label{fig:figure05}
\end{figure*}

The local parallel-plate depletion adhesion density is
\begin{equation}
w_{\mathrm{dep}}=2\Pi r.
\label{eq:wdep}
\end{equation}
Two parallel planes in contact exclude crowder centers from a layer of thickness $r$ at each plane; the two excluded layers overlap over a length $2r$ per unit area, so the free-energy decrease per unit area is $2\Pi r$. With CS pressure,
\begin{equation}
w_{\mathrm{dep}}^{\mathrm{CS}}=2ck_{\mathrm{B}}TrZ_{\mathrm{CS}}(\phi).
\label{eq:wdepCS}
\end{equation}
At the representative point $r=15\,\mathrm{nm}$,
\begin{equation}
w_{\mathrm{dep}}^{\mathrm{CS}}=4.75303\times10^{-6}\,\mathrm{J\,m^{-2}},
\label{eq:wdeprep}
\end{equation}
which is about $1.53\%$ of the representative total adhesion $w_{\mathrm{total}}=3.1\times10^{-4}\,\mathrm{J\,m^{-2}}$. At $r=20\,\mathrm{nm}$, the ratio can exceed $10\%$, and the state warning is strongest there. Figure~\ref{fig:figure04} displays $Z_{\mathrm{CS}}(\phi)$ and $w_{\mathrm{dep}}$ versus $\phi$, and Fig.~\ref{fig:figure05} shows the mapping from $r$ to $\phi$ and to the osmotic pressure at fixed Table-I number density.

\subsection{Deterministic approach with bulk Stokes drag}
\label{subsec:approach-bulk}

The capture distance is denoted by $\delta$. The ideal AO exclusion layers begin to overlap at $d=2r$. A deterministic trajectory started exactly at $d=2r$ experiences zero force. A finite entry offset $d_0=2r-\epsilon$ with $\epsilon>0$ is used for the deterministic approach estimate. In the numerical estimates below we use $\epsilon=0.01\,\mathrm{nm}$, subject to $0<\epsilon<2r-\delta$; because the approach time diverges logarithmically as $\epsilon\to0$, the sub-millisecond estimate is conditional on this chosen entry offset and on the chosen pressure closure. With bulk Stokes drag $6\pi\mu R$ and force \eqref{eq:FAOmag}, the overdamped balance is
\begin{equation}
6\pi\mu R(-\dot d)=\pi\Pi(2r-d)(2R+d).
\label{eq:balance}
\end{equation}
The time from $d_0=2r-\epsilon$ to $d=\delta$ is
\begin{equation}
t_{\mathrm{bulk}}
=\frac{6\mu R}{\Pi}\int_{\delta}^{2r-\epsilon}
\frac{\mathrm{d}d}{(2r-d)(2R+d)}.
\label{eq:tbulkint}
\end{equation}
Using
\begin{equation}
\frac{1}{(2r-d)(2R+d)}
=\frac{1}{2(R+r)}\left[\frac{1}{2r-d}+\frac{1}{2R+d}\right],
\label{eq:partialfrac}
\end{equation}
the integral gives
\begin{equation}
t_{\mathrm{bulk}}
=\frac{3\mu R}{\Pi(R+r)}
\ln\left[
\frac{(2r-\delta)(2R+2r-\epsilon)}
{\epsilon(2R+\delta)}
\right].
\label{eq:tbulk}
\end{equation}
Near $d=2r$, writing $s=2r-d$ gives
\begin{equation}
|F_{\mathrm{AO}}|
\sim 2\pi\Pi(R+r)s.
\label{eq:Fnear}
\end{equation}
The time integral contains $\int_\epsilon \mathrm{d}s/s=\ln(1/\epsilon)$, so a deterministic trajectory started exactly at $d=2r$ has a logarithmically divergent entry time. Thermal fluctuations are required to enter the depletion layer from $d=2r$. The finite offset $\epsilon$ separates the thermal entry event from the deterministic drift after entry. The time \eqref{eq:tbulk} is conditional on the particle having entered the depletion layer. The stochastic entry and first-bond formation that precede this deterministic drift are treated in the SR theory of Ref.~\cite{wu2026single}.

\subsection{Brenner normal wall drag}
\label{subsec:brenner}

Near a rigid no-slip wall, the normal drag is larger than the bulk Stokes value. Write
\begin{equation}
F_{\mathrm{drag}}^{\perp}=6\pi\mu R\lambda_{\perp}(d/R)v,
\label{eq:dragB}
\end{equation}
where $v=-\dot d>0$. The Brenner normal drag factor is \cite{brenner1961}
\begin{widetext}
\begin{equation}
\lambda_{\perp}
=\frac{4}{3}\sinh\alpha
\sum_{n=1}^{\infty}
\frac{n(n+1)}{(2n-1)(2n+3)}
\left[
\frac{2\sinh[(2n+1)\alpha]+(2n+1)\sinh(2\alpha)}
{4\sinh^2[(n+1/2)\alpha]-(2n+1)^2\sinh^2\alpha}
-1
\right],
\label{eq:brenner}
\end{equation}
\end{widetext}
with
\begin{equation}
\cosh\alpha=1+\frac{d}{R}.
\label{eq:alpha}
\end{equation}
The force balance becomes
\begin{equation}
6\pi\mu R\lambda_{\perp}(d/R)(-\dot d)=\pi\Pi(2r-d)(2R+d),
\label{eq:balanceB}
\end{equation}
and the wall-corrected approach time is
\begin{equation}
t_{\mathrm{B}}
=\frac{6\mu R}{\Pi}
\int_{\delta}^{2r-\epsilon}
\frac{\lambda_{\perp}(d/R)}{(2r-d)(2R+d)}\,\mathrm{d}d.
\label{eq:tB}
\end{equation}
Because $\lambda_{\perp}>1$, one has $t_{\mathrm{B}}>t_{\mathrm{bulk}}$ for the same geometry and pressure. In the Table-I scan, the ratio $t_{\mathrm{B}}/t_{\mathrm{bulk}}$ ranges approximately from $2.8$ to $12.6$, depending on $R$ and $r$. The capture time lies in the sub-millisecond range in the deterministic estimate for the chosen $\epsilon=0.01\,\mathrm{nm}$ and CS pressure, whereas the wrapping first-passage time is of order $0.5$ to $1\,\mathrm{s}$ for the representative parameters. This comparison is conditional; it does not include the stochastic first entry into the depletion layer, receptor binding waiting times, or a full stochastic total endocytosis time~\cite{wu2026single}.

\section{Wrapping geometry and free energy}
\label{sec:wrapping-free-energy}

\subsection{Spherical-cap coordinate}
\label{subsec:cap-coordinate}

Let the membrane in the attached region follow the particle sphere of radius $R$. Let the contact line correspond to polar angle $\theta$. The spherical-cap height from the pole to the contact plane is $h=R(1-\cos\theta)$. Define the dimensionless wrapping coordinate
\begin{equation}
x=\frac{h}{R}=1-\cos\theta,\quad 0\le\theta\le\pi,\quad 0\le x\le2.
\label{eq:xdef}
\end{equation}
The attached spherical area is
\begin{equation}
A_c=2\pi R^2(1-\cos\theta)=2\pi R^2x,
\label{eq:Ac}
\end{equation}
so the wrapped area fraction is
\begin{equation}
f_A=\frac{A_c}{4\pi R^2}=\frac{x}{2}.
\label{eq:fA}
\end{equation}
The $95\%$ spherical-area criterion corresponds to $f_A=0.95$, $x_f=1.9$, leaving only a small unwrapped cap of height $0.1R$; complete wrapping corresponds to $f_A=1$ and $x=2$. Thus $x_f=1.9$ describes a deep, nearly closed wrapped state, and we use it as an operational near-complete threshold rather than as a topological completion criterion. The geometrical details of this spherical-cap parametrization are collected in Appendix~\hyperref[app:cap]{B}.

\subsection{Helfrich bending energy}
\label{subsec:bending}

For zero spontaneous curvature and with the Gaussian boundary contribution absorbed into the reduced description, the Helfrich bending energy on the attached spherical cap is \cite{helfrich1973,Canham1970,Ouyang1989,wu2018getting,wu2023,wu2025generalized}
\begin{equation}
G_{\mathrm{bend}}
=\int_{A_c}\frac{\kappa}{2}(2H)^2\,\mathrm{d}A.
\label{eq:Gbendint}
\end{equation}
For a sphere of radius $R$, $H=1/R$ and $2H=2/R$, so the bending energy density is $2\kappa/R^2$. Multiplying by $A_c=2\pi R^2x$ gives
\begin{equation}
G_{\mathrm{bend}}
=4\pi\kappa x.
\label{eq:Gbend}
\end{equation}
The bending energy is linear in $x$. The Gaussian curvature contribution of an open cap is not constant by itself. The reduced model assumes that the total Gaussian contribution of the attached region, the outer membrane, and the contact-line boundary cancels between compared states or is absorbed into contact-line and constant terms.

\subsection{Membrane tension energy}
\label{subsec:tension}

The contact circle projected onto the undeformed membrane has radius $a_{\mathrm{wrap}}=R\sin\theta$. From $\cos\theta=1-x$,
\begin{equation}
a_{\mathrm{wrap}}=R\sqrt{2x-x^2},
\label{eq:awrap}
\end{equation}
so the projected area is
\begin{equation}
A_p=\pi a_{\mathrm{wrap}}^2=\pi R^2(2x-x^2).
\label{eq:Ap}
\end{equation}
The excess membrane area relative to the projected disk is
\begin{equation}
\Delta A=A_c-A_p
=\pi R^2x^2.
\label{eq:DeltaA}
\end{equation}
For constant membrane tension $\gamma$, the tension energy is
\begin{equation}
G_{\mathrm{ten}}=\gamma\Delta A=\pi\gamma R^2x^2.
\label{eq:Gten}
\end{equation}
This quadratic term assumes a constant tension reservoir and a flat outer membrane. It does not resolve the transition region outside the contact line or a neck. Membrane tension also plays a dynamic role in the flat-to-curved transition during clathrin-mediated endocytosis \cite{BucherEtAl2018}, and adhesion energy controls lipid binding-mediated endocytosis \cite{GrozaEtAl2024}.

\subsection{Adhesion energy}
\label{subsec:adhesion}

Let $w_{\mathrm{LR}}>0$ be the ligand-receptor adhesion energy density. In the local parallel approximation for $r/R\ll1$, the depletion adhesion density is $w_{\mathrm{dep}}=2\Pi r>0$. Define the total adhesion density
\begin{equation}
w_{\mathrm{total}}=w_{\mathrm{LR}}+w_{\mathrm{dep}}.
\label{eq:wtotal}
\end{equation}
Attachment lowers the free energy, so
\begin{equation}
G_{\mathrm{adh}}=-w_{\mathrm{total}}A_c=-2\pi R^2w_{\mathrm{total}}x.
\label{eq:Gadh}
\end{equation}
The model uses either a total adhesion input $w_{\mathrm{total}}$ that already includes depletion or a ligand adhesion input $w_{\mathrm{LR}}$ to which $w_{\mathrm{dep}}$ is added. The two inputs are mutually exclusive in the numerical implementation to avoid double counting. The local parallel-plate depletion adhesion is a reduced approximation for the wrapped state. It is not the same geometric integral as the sphere-plane contact energy $U_{\mathrm{AO}}(0)$. At the representative point $r/R=15/50=0.3$, which is not very small, so this local parallel-plate density is only a first approximation; the exact AO sphere-plane geometry is used only in the capture stage. Recent studies have examined ligand mobility and configurational entropy in nanoparticle uptake \cite{ZhangLiWang2021PRE,ChenXueLiEtAl2024}, and steric interactions between mobile ligands can facilitate complete wrapping in passive endocytosis \cite{DiMicheleJanaMognetti2018}. These effects are not explicitly resolved in the continuum binding energy density; they influence the effective value of $w_{\mathrm{total}}$. The multivalent bond stabilization after the first bond, which can further modify the effective adhesion, is treated separately in Ref.~\cite{wu2026multivalent}.

\begin{figure*}[t]
\centering
\includegraphics[width=\linewidth]{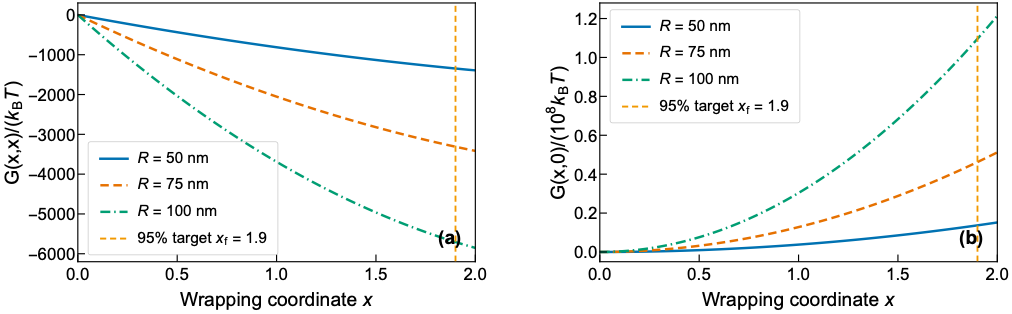}
\caption{Quadratic single-mode free energy at total adhesion $w_{\mathrm{total}}=3.1\times10^{-4}\,\mathrm{J\,m^{-2}}$ for $R=50$, $75$, and $100\,\mathrm{nm}$. (a) Relaxed slice $q=x$ showing the long-time landscape; the target $x_f=1.9$ is marked; the vertical axis is in units of $k_BT$. (b) Instantaneous slice $q=0$ showing the much larger initial stiffness; the vertical axis is in units of $10^8 k_BT$, and this slice is a fixed-internal-variable energy landscape, not the energy along the actual wrapping trajectory.}
\label{fig:figure06}
\end{figure*}

\subsection{Single-mode standard linear solid and total free energy}
\label{subsec:sls-free-energy}

The single-mode standard linear solid is represented by an equilibrium spring $E_\infty$ in parallel with a Maxwell branch of spring modulus $E_1$ and dashpot viscosity $\eta_M$. The internal strain variable is $q$. The cortical free energy is
\begin{equation}
G_{\mathrm{cyt}}(x,q)=\frac{\chi R^3}{2}\left[E_\infty x^2+E_1(x-q)^2\right].
\label{eq:Gcyt}
\end{equation}
The dimensionless geometry factor $\chi>0$ collects the shape-function, Poisson-ratio, boundary-condition, and tensor-contraction integrals of a single-mode displacement scaling. The internal-variable dissipation potential is
\begin{equation}
\Phi_q=\frac{1}{2}C\dot q^2,\qquad C=\chi\eta_MR^3,
\label{eq:Phiq}
\end{equation}
and the wrapping-coordinate dissipation is
\begin{equation}
\Phi_x=\frac{1}{2}\Xi\dot x^2,\qquad \Xi=6\pi\mu R^3.
\label{eq:Phix}
\end{equation}
The coefficient $\Xi$ is a single-mode effective drag obtained from the Stokes scale $6\pi\mu R$ and the displacement scale $R$; it is not a nonlocal kernel obtained from a full membrane--fluid boundary-value problem.

Combining \eqref{eq:Gbend}, \eqref{eq:Gten}, \eqref{eq:Gadh}, and \eqref{eq:Gcyt} gives
\begin{equation}
\begin{aligned}
G(x,q)
=&4\pi\kappa x+\pi\gamma R^2x^2-2\pi R^2w_{\mathrm{total}}x \\
&+\frac{\chi R^3}{2}\left[E_\infty x^2+E_1(x-q)^2\right].
\label{eq:Gexpanded}
\end{aligned}
\end{equation}
Define the energy coefficients
\begin{equation}
\begin{aligned}
A=2\pi R^2w_{\mathrm{total}}-4\pi\kappa,
\quad
K_\gamma=2\pi\gamma R^2,
\\
K_\infty=\chi E_\infty R^3,
\quad
K_1=\chi E_1 R^3.
\label{eq:coeffs}
\end{aligned}
\end{equation}
Collecting linear and quadratic terms gives the compact form
\begin{equation}
G(x,q)=-Ax+\frac{1}{2}(K_\gamma+K_\infty)x^2+\frac{1}{2}K_1(x-q)^2.
\label{eq:Gcompact}
\end{equation}
The coefficient $A$ is a linear driving coefficient, not a positive-definite stiffness. The Hessian of $G$ with respect to $(x,q)$ is
\begin{equation}
H_G=
\begin{pmatrix}
K_\gamma+K_\infty+K_1 & -K_1\\
-K_1 & K_1
\end{pmatrix}.
\label{eq:Hessian}
\end{equation}
For $K_1>0$ and $K_\gamma+K_\infty>0$, the first leading principal minor is positive and
\begin{equation}
\det H_G=K_1(K_\gamma+K_\infty)>0.
\label{eq:Hdet}
\end{equation}
By Sylvester's criterion, the Hessian $H_G$ is positive definite, and the Hessian convexity of $G$ on $\mathbb{R}^2$ follows. This Hessian convexity is what later guarantees a unique equilibrium and a globally asymptotically stable unconstrained gradient flow. The unconstrained minimum satisfies $\partial_xG=\partial_qG=0$. The second equation gives $q_\infty=x_\infty$, and the first gives
\begin{equation}
x_\infty=\frac{A}{K_\gamma+K_\infty}
=\frac{2\pi R^2w_{\mathrm{total}}-4\pi\kappa}{2\pi\gamma R^2+\chi E_\infty R^3}.
\label{eq:xinf}
\end{equation}
The modulus $E_1$ does not enter $x_\infty$; it affects the transient time scales. The relaxed ($q=x$) and instantaneous ($q=0$) free-energy slices are shown in Fig.~\ref{fig:figure06}.

\section{Exact mapping of the creep compliance to the internal-variable standard linear solid}
\label{sec:sls-mapping}

\begin{figure*}[t]
\centering
\includegraphics[width=\linewidth]{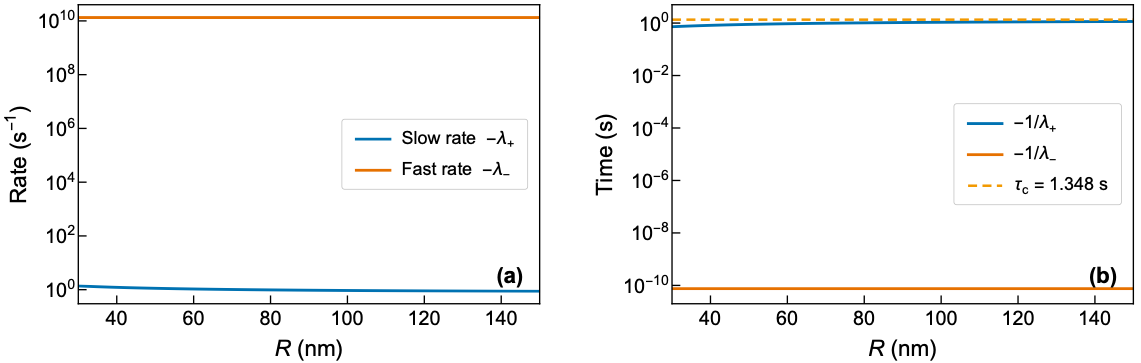}
\caption{Relaxation spectrum of the two-variable Onsager system for $R=30$--$150\,\mathrm{nm}$. (a) Slow rate $-\lambda_+$ and fast rate $-\lambda_-$ versus $R$ on a logarithmic scale; the fast mode is an external-drag/instantaneous-stiffness mode, nearly flat at $\sim10^{10}\,\mathrm{s^{-1}}$, while the slow mode is of order $1\,\mathrm{s^{-1}}$. The physical applicability of the extremely short fast time is not established. (b) Corresponding time scales $-1/\lambda_{\pm}$ versus $R$; the slow time is $\sim0.7$--$1.15\,\mathrm{s}$, distinct from the paper creep retardation time $\tau_c=1.348\,\mathrm{s}$ (dashed).}
\label{fig:figure13}
\end{figure*}

Our recent theoretical framework gives the creep compliance in Eq.~(26) of \cite{Liu2026e2} as
\begin{equation}
J(t)=\frac{1}{E_v}+\frac{1}{E_c}\left(1-e^{-t/\tau_c}\right),
\quad
\tau_c=\frac{\eta_c}{E_c}.
\label{eq:Jpaper}
\end{equation}
The instantaneous and long-time compliances are $J(0^+)=1/E_v$ and $J(\infty)=1/E_v+1/E_c$, so the corresponding moduli are
\begin{equation}
E_0=\frac{1}{J(0^+)}=E_v,
\quad
E_\infty=\frac{1}{J(\infty)}=\frac{E_vE_c}{E_v+E_c}.
\label{eq:E0Einf}
\end{equation}
The Maxwell spring modulus is
\begin{equation}
E_1=E_0-E_\infty
=\frac{E_v^2}{E_v+E_c}.
\label{eq:E1}
\end{equation}

For the internal-variable model, take the unit-volume free energy
\begin{equation}
\psi(x,q)=\frac{1}{2}E_\infty x^2+\frac{1}{2}E_1(x-q)^2,
\label{eq:psi}
\end{equation}
and the dashpot dissipation density $\phi_q(\dot q)=\eta_M\dot q^2/2$. The stress is
\begin{equation}
\sigma=\frac{\partial\psi}{\partial x}
=E_\infty x+E_1(x-q)=E_0x-E_1q.
\label{eq:sigma}
\end{equation}
Stationarity with respect to $\dot q$ gives
\begin{equation}
\eta_M\dot q=-\frac{\partial\psi}{\partial q}=E_1(x-q).
\label{eq:qode}
\end{equation}
Differentiating \eqref{eq:sigma} and eliminating $q$ with \eqref{eq:qode} gives the constitutive equation
\begin{equation}
\dot\sigma+\frac{E_1}{\eta_M}\sigma
=E_0\dot x+\frac{E_1E_\infty}{\eta_M}x.
\label{eq:constitutive1}
\end{equation}
Define the stress relaxation time
\begin{equation}
\tau_{\mathrm{rel}}=\frac{\eta_M}{E_1}.
\label{eq:taurel}
\end{equation}
Multiplying \eqref{eq:constitutive1} by $\tau_{\mathrm{rel}}$ gives
\begin{equation}
\sigma+\tau_{\mathrm{rel}}\dot\sigma
=E_\infty x+E_0\tau_{\mathrm{rel}}\dot x.
\label{eq:diffcon}
\end{equation}

For a step stress $\sigma(t)=\sigma_0H(t)$, $\dot\sigma=0$ for $t>0$, and \eqref{eq:constitutive1} becomes
\begin{equation}
\dot x+\frac{x}{\tau_c}=\frac{\sigma_0}{E_\infty\tau_c},
\quad
\tau_c=\frac{\eta_ME_0}{E_1E_\infty}.
\label{eq:stepstress3}
\end{equation}
At the step instant the dashpot length cannot jump, so $q(0^+)=0$ and $x(0^+)=\sigma_0/E_0$. The solution is
\begin{equation}
x(t)=\frac{\sigma_0}{E_\infty}
+\left(\frac{\sigma_0}{E_0}-\frac{\sigma_0}{E_\infty}\right)e^{-t/\tau_c},
\label{eq:xstep}
\end{equation}
so the creep compliance of the internal-variable model is
\begin{equation}
J_{\mathrm{SLS}}(t)
=\frac{1}{E_\infty}
-\left(\frac{1}{E_\infty}-\frac{1}{E_0}\right)e^{-t/\tau_c}.
\label{eq:JSLS}
\end{equation}
Matching \eqref{eq:JSLS} with \eqref{eq:Jpaper} gives the endpoint relations and the time relation. The constant terms give $1/E_\infty=1/E_v+1/E_c$, which yields \eqref{eq:E0Einf}. The exponential coefficients give
\begin{equation}
\frac{1}{E_\infty}-\frac{1}{E_0}=\frac{1}{E_c},
\label{eq:matchexp}
\end{equation}
which together with \eqref{eq:E0Einf} gives $E_0=E_v$. Matching the creep retardation time $\tau_c=\eta_c/E_c$ with \eqref{eq:stepstress3} gives
\begin{equation}
\eta_M=\frac{\eta_c}{E_c}\frac{E_1E_\infty}{E_0}
=\eta_c\frac{E_v^2}{(E_v+E_c)^2}.
\label{eq:etaMexplicit}
\end{equation}
The stress relaxation time is
\begin{equation}
\tau_{\mathrm{rel}}
=\frac{\eta_M}{E_1}
=\frac{\eta_c}{E_v+E_c}.
\label{eq:taurelax}
\end{equation}
The creep retardation time and the stress relaxation time are distinct. In the Table-I representative values,
\begin{equation}
\tau_c=1.348\,\mathrm{s},
\qquad
\tau_{\mathrm{rel}}=2.695946\times10^{-5}\,\mathrm{s}.
\end{equation}
The slow time of the coupled two-variable system is not equal to either isolated branch time; it is determined by the full matrix spectrum. The complete algebraic mapping is given in Appendix~\hyperref[app:sls]{C}, and the derived single-mode standard linear solid parameters are listed in Table~\ref{tab:sls}.

For a single-mode displacement scaling, the cortical free energy and dissipation become
\begin{equation}
\begin{aligned}
G_{\mathrm{cyt}}(x,q)=\frac{\chi R^3}{2}\left[E_\infty x^2+E_1(x-q)^2\right],
\\
\Phi_q=\frac{1}{2}\chi\eta_MR^3\dot q^2.
\label{eq:Gcyt2}
\end{aligned}
\end{equation}
With $K_\infty=\chi E_\infty R^3$, $K_1=\chi E_1 R^3$, and $C=\chi\eta_MR^3$, the internal-variable equation is
\begin{equation}
C\dot q=K_1(x-q).
\label{eq:qode2}
\end{equation}
The isolated internal time constant is
\begin{equation}
\frac{C}{K_1}
=\frac{\eta_M}{E_1}
=\tau_{\mathrm{rel}}.
\label{eq:internal-time}
\end{equation}

The geometry factor $\chi$ is fixed by matching the long-time static elastic coefficient at complete wrapping $x_f=2$. The static elastic threshold coefficient is given in Eq.~(52) of \cite{Liu2026e2} as
\begin{equation}
w_{\mathrm{cyt,crit}}^{\mathrm{paper}}
=\frac{16\sqrt{2}}{9\pi}E_\infty R.
\label{eq:paperstatic}
\end{equation}
The quadratic single-mode model gives the elastic contribution at target $x_f$ as
\begin{equation}
w_{\mathrm{cyt,crit}}
=\frac{\chi E_\infty R x_f}{2\pi}.
\label{eq:modelstatic}
\end{equation}
Setting $x_f=2$ and comparing gives
\begin{equation}
\chi=\frac{16\sqrt{2}}{9}=2.514157444.
\label{eq:chi}
\end{equation}
This matching fixes a long-time static coefficient. It does not retain the Hertz $h^{3/2}$ nonlinear contact law used in Eqs.~(29)--(41) of \cite{Liu2026e2}. The quadratic free energy is used because it gives a strictly convex two-variable linear gradient flow with an explicit spectrum and a unique first-passage root. The two relaxation rates and associated time scales are shown in Fig.~\ref{fig:figure13}.

\begin{table}[t]
\centering
\caption{Derived single-mode standard linear solid parameters at the Table-I representative values. The stress relaxation time is $\eta_c/(E_v+E_c)$. The creep retardation time is $\eta_c/E_c$.}
\label{tab:sls}
\begin{tabular}{lll}
\toprule
Quantity & Value & Unit\\
\midrule
$E_0=E_v$ & $1.000000000\times10^8$ & $\mathrm{Pa}$\\
$E_\infty=E_vE_c/(E_v+E_c)$ & $1999.9600008$ & $\mathrm{Pa}$\\
$E_1=E_0-E_\infty$ & $9.999800004\times10^7$ & $\mathrm{Pa}$\\
$\tau_c=\eta_c/E_c$ & $1.348$ & $\mathrm{s}$\\
$\eta_M=\eta_cE_v^2/(E_v+E_c)^2$ & $2695.8921632$ & $\mathrm{Pa\,s}$\\
$\tau_{\mathrm{rel}}=\eta_M/E_1$ & $2.69594608\times10^{-5}$ & $\mathrm{s}$\\
$\chi=16\sqrt{2}/9$ & $2.514157444$ & dimensionless\\
$x_f=1.9$ & $1.9$ & dimensionless\\
\bottomrule
\end{tabular}
\end{table}

\section{Two-variable RO gradient flow}
\label{sec:gradient-flow}

\subsection{Dissipation and variational equations}
\label{subsec:variational}

The total dissipation potential is
\begin{equation}
\Phi(\dot x,\dot q)=\frac{1}{2}\Xi\dot x^2+\frac{1}{2}C\dot q^2,
\quad \Xi>0,\quad C>0.
\label{eq:Phi}
\end{equation}
The Rayleighian is
\begin{equation}
\mathcal{R}
=\frac{\partial G}{\partial x}\dot x+\frac{\partial G}{\partial q}\dot q
+\frac{1}{2}\Xi\dot x^2+\frac{1}{2}C\dot q^2.
\label{eq:Rayleighian}
\end{equation}
Stationarity with respect to the rates gives the Onsager equations
\begin{equation}
\Xi\dot x=-\frac{\partial G}{\partial x},
\quad
C\dot q=-\frac{\partial G}{\partial q}.
\label{eq:onsager}
\end{equation}
From \eqref{eq:Gcompact},
\begin{equation}
\begin{aligned}
\frac{\partial G}{\partial x}
&=-A+(K_\gamma+K_\infty)x+K_1(x-q),
\\
\frac{\partial G}{\partial q}&=-K_1(x-q).
\label{eq:GxGq}
\end{aligned}
\end{equation}
Substitution gives
\begin{equation}
\Xi\dot x
=A-(K_\gamma+K_\infty+K_1)x+K_1q,
\label{eq:xdot}
\end{equation}
and
\begin{equation}
C\dot q=K_1(x-q).
\label{eq:qdot}
\end{equation}
Multiplying \eqref{eq:xdot} by $\dot x$ and \eqref{eq:qdot} by $\dot q$ and adding gives the Lyapunov identity
\begin{equation}
\frac{\mathrm{d}G}{\mathrm{d}t}
=\frac{\partial G}{\partial x}\dot x+\frac{\partial G}{\partial q}\dot q
=-\Xi\dot x^2-C\dot q^2\le0.
\label{eq:energy}
\end{equation}
Equivalently, $G$ is a Lyapunov function for the unconstrained gradient flow. The equality holds only when $\dot x=\dot q=0$. Together with the Hessian convexity of $G$, the Lyapunov identity implies that the unconstrained equilibrium is unique and globally asymptotically stable.

\subsection{Matrix form and spectrum}
\label{subsec:spectrum}

\begin{figure*}[t]
\centering
\includegraphics[width=\linewidth]{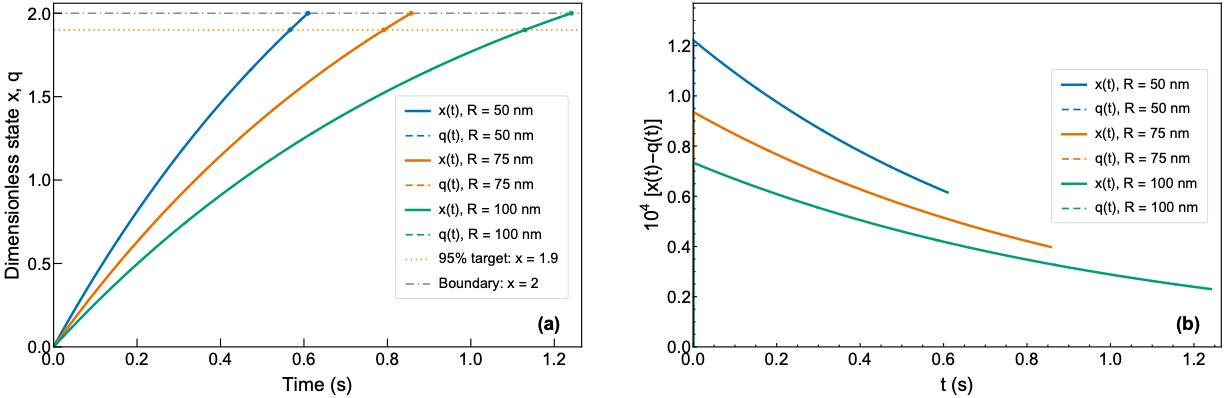}
\caption{Exact two-exponential interior trajectories at total adhesion $w_{\mathrm{total}}=3.1\times10^{-4}\,\mathrm{J\,m^{-2}}$ for $R=50$, $75$, and $100\,\mathrm{nm}$. (a) Wrapping coordinate $x(t)$ and internal variable $q(t)$ up to the first hit of the physical boundary $x=2$; circles mark $x_f=1.9$ and squares mark the first boundary hit. (b) Magnified difference $x(t)-q(t)$ (scaled by $10^4$), showing that $q$ follows $x$ closely after the rapid internal relaxation while retaining the small lag that carries the SLS memory.}
\label{fig:figure07}
\end{figure*}

Define
\begin{equation}
\begin{aligned}
a=\frac{K_\gamma+K_\infty+K_1}{\Xi},
\qquad
b=\frac{K_1}{\Xi},
\\
c=\frac{K_1}{C}=\frac{1}{\tau_{\mathrm{rel}}},
\qquad
d_0=\frac{A}{\Xi}.
\label{eq:abc}
\end{aligned}
\end{equation}
Then \eqref{eq:xdot} and \eqref{eq:qdot} become
\begin{equation}
\dot x=-ax+bq+d_0,
\quad
\dot q=cx-cq.
\end{equation}
In matrix form,
\begin{equation}
\frac{\mathrm{d}}{\mathrm{d}t}
\begin{pmatrix}
x\\ q
\end{pmatrix}
=
\begin{pmatrix}
-a & b\\
c & -c
\end{pmatrix}
\begin{pmatrix}
x\\ q
\end{pmatrix}
+
\begin{pmatrix}
d_0\\ 0
\end{pmatrix}.
\label{eq:matrix}
\end{equation}
Since all stiffnesses and dampings are positive,
\begin{equation}
a-b
=\frac{K_\gamma+K_\infty}{\Xi}>0.
\label{eq:ab}
\end{equation}
The characteristic polynomial is
\begin{equation}
\det(\lambda I-M)
=\lambda^2+(a+c)\lambda+c(a-b).
\label{eq:charpoly}
\end{equation}
The discriminant is
\begin{equation}
\Delta=(a+c)^2-4c(a-b)
=(a-c)^2+4bc>0.
\label{eq:disc}
\end{equation}
The two eigenvalues are real and distinct:
\begin{equation}
\lambda_\pm=-\frac{a+c}{2}\pm\frac{1}{2}\sqrt{(a-c)^2+4bc}.
\label{eq:eigen}
\end{equation}
The sum and product satisfy
\begin{equation}
\lambda_++\lambda_-=-(a+c)<0,
\quad
\lambda_+\lambda_-=c(a-b)>0,
\label{eq:eigensumprod}
\end{equation}
so both roots have negative real parts. Since the roots are real,
\begin{equation}
\lambda_-<\lambda_+<0.
\label{eq:ordering}
\end{equation}
The system has two purely relaxing modes and no oscillatory mode. The matrix spectrum and velocity residues are derived in Appendix~\hyperref[app:matrix]{D}.

\subsection{Exact double-exponential solution}
\label{subsec:exact-solution}

The steady state is
\begin{equation}
x_\infty=q_\infty=\frac{d_0}{a-b}
=\frac{A}{K_\gamma+K_\infty}.
\end{equation}
For $x(0)=q(0)=0$, the wrapping coordinate is
\begin{equation}
x(t)=x_\infty+C_+e^{\lambda_+t}+C_-e^{\lambda_-t}.
\label{eq:xt}
\end{equation}
The initial condition gives $C_++C_-=-x_\infty$, and the initial velocity $\dot x(0)=d_0$ gives
\begin{equation}
\lambda_+C_++\lambda_-C_-=\frac{A}{\Xi}.
\label{eq:Clambda}
\end{equation}
Solving the two linear equations gives
\begin{equation}
C_+=\frac{d_0+\lambda_-x_\infty}{\lambda_+-\lambda_-},
\quad
C_-=\frac{-d_0-\lambda_+x_\infty}{\lambda_+-\lambda_-}.
\label{eq:Cplusminus}
\end{equation}

A numerically stable representation uses the positive residues of the velocity. Write
\begin{equation}
\dot x(t)=r_+e^{\lambda_+t}+r_-e^{\lambda_-t},
\label{eq:xdotres}
\end{equation}
with $r_+=\lambda_+C_+$ and $r_-=\lambda_-C_-$. Using $\lambda_+\lambda_-=c(a-b)$ and $x_\infty=d_0/(a-b)$,
\begin{equation}
r_+=\frac{A}{\Xi}\frac{\lambda_++c}{\lambda_+-\lambda_-},
\quad
r_-=\frac{A}{\Xi}\frac{\lambda_-+c}{\lambda_--\lambda_+}.
\label{eq:rplusminus}
\end{equation}
Integration of \eqref{eq:xdotres} with $x(0)=0$ gives
\begin{equation}
x(t)=r_+\frac{e^{\lambda_+t}-1}{\lambda_+}
+r_-\frac{e^{\lambda_-t}-1}{\lambda_-}.
\label{eq:xtres}
\end{equation}
The internal variable can be reconstructed from \eqref{eq:xdot} as
\begin{equation}
q(t)=\frac{\Xi\dot x(t)-A+(K_\gamma+K_\infty+K_1)x(t)}{K_1}.
\label{eq:qt}
\end{equation}
This expression contains only decaying exponentials and avoids forming products of large intermediate exponentials. The exact $x(t)$ and $q(t)$ trajectories up to the first hit of $x=2$ are shown in Fig.~\ref{fig:figure07}.

\subsection{Monotonicity from the Metzler positive semigroup}
\label{subsec:metzler}

The matrix
\begin{equation}
M=
\begin{pmatrix}
-a & b\\
c & -c
\end{pmatrix}
\label{eq:M}
\end{equation}
has nonnegative off-diagonal entries $b>0$ and $c>0$. It is an irreducible Metzler matrix. For any Metzler matrix, $e^{Mt}$ is entrywise nonnegative for $t\ge0$; for an irreducible Metzler matrix, $e^{Mt}$ is entrywise positive for $t>0$. Let $v(t)=\dot y(t)$ with $y=(x,q)^T$. Differentiating \eqref{eq:matrix} gives
\begin{equation}
\dot v=Mv,
\quad
v(0)=
\begin{pmatrix}
A/\Xi\\ 0
\end{pmatrix}.
\label{eq:v0}
\end{equation}
For $A>0$, $v(t)=e^{Mt}v(0)$ has two positive components for $t>0$. In particular,
\begin{equation}
\dot x(t)>0,\quad t>0.
\label{eq:monotoneM}
\end{equation}
The same argument gives $\dot q(t)>0$ for $t>0$.

\subsection{Monotonicity from Laplace residues}
\label{subsec:laplace}

Eliminating $q$ from \eqref{eq:xdot} and \eqref{eq:qdot} gives the scalar equation
\begin{equation}
\ddot x+(a+c)\dot x+c(a-b)x=cd_0.
\label{eq:scalar}
\end{equation}
With $x(0)=0$ and $\dot x(0)=d_0$, the Laplace transform of \eqref{eq:scalar} gives
\begin{equation}
sX(s)=\frac{A}{\Xi}\frac{s+c}{(s-\lambda_+)(s-\lambda_-)}.
\label{eq:laplace}
\end{equation}
The characteristic polynomial evaluated at $s=-c$ is
\begin{equation}
P(-c)=-bc<0.
\label{eq:Pc}
\end{equation}
Because the quadratic opens upward and has two real roots, the zero $-c$ lies between the two poles:
\begin{equation}
\lambda_-<-c<\lambda_+<0.
\label{eq:interlace}
\end{equation}
The partial-fraction expansion is
\begin{equation}
\dot x(t)=\frac{A}{\Xi}
\left[
\frac{\lambda_++c}{\lambda_+-\lambda_-}e^{\lambda_+t}
-\frac{\lambda_-+c}{\lambda_+-\lambda_-}e^{\lambda_-t}
\right].
\label{eq:xdotpartial}
\end{equation}
From \eqref{eq:interlace}, $\lambda_++c>0$ and $\lambda_-+c<0$, so the two coefficients in \eqref{eq:xdotpartial} are positive. For $A>0$,
\begin{equation}
\dot x(t)>0,\quad t>0.
\label{eq:monotoneL}
\end{equation}
The two arguments in \eqref{eq:monotoneM} and \eqref{eq:monotoneL} are independent. Negative eigenvalues alone do not imply componentwise monotonicity; the Metzler structure and the residue signs supply the additional information.

\subsection{First-passage equation and uniqueness}
\label{subsec:first-passage}

For
\begin{equation}
0<x_f<x_\infty,
\label{eq:fpcondition}
\end{equation}
one has $x(0)=0<x_f$, $x(t)$ is continuous and strictly increasing, and $x(t)\to x_\infty>x_f$. The intermediate value theorem gives at least one $t_f>0$ with
\begin{equation}
x(t_f)=x_f.
\label{eq:fpeq}
\end{equation}
Strict monotonicity gives uniqueness. The first-passage time is obtained from the scalar equation \eqref{eq:fpeq}. The two noncommensurate exponentials generally do not admit an elementary analytic inverse, so a bracketing root solver is used on the unique scalar root.

Near the reachability boundary, let $\delta=x_\infty-x_f>0$. The long-time approach is controlled by the slow mode. Write
\begin{equation}
x_\infty-x(t)=B_+e^{\lambda_+t}+B_-e^{\lambda_-t},
\quad B_+>0.
\end{equation}
When $t_f$ is large, the fast mode is negligible and $\delta\simeq B_+e^{\lambda_+t_f}$, so
\begin{equation}
t_f\simeq\frac{1}{|\lambda_+|}\ln\frac{B_+}{\delta}.
\label{eq:logdiv}
\end{equation}
The first-passage time diverges logarithmically as the target approaches the long-time limit.

\section{Projected dynamics on the physical interval}
\label{sec:projected}

The spherical-cap coordinate satisfies $0\le x\le2$. The unconstrained linear system is defined on $\mathbb{R}^2$. When $A<0$, the unconstrained initial velocity $\dot x(0)=A/\Xi<0$ points outside the physical interval. When $x_\infty>2$, the unconstrained trajectory crosses $x=2$ at a finite time. The projected Onsager dynamics uses the normal cone to $[0,2]$. The tangent cone is
\begin{equation}
T_{[0,2]}(x)=
\begin{cases}
[0,\infty), & x=0,\\
\mathbb{R}, & 0<x<2,\\
(-\infty,0], & x=2,
\end{cases}
\label{eq:tangent}
\end{equation}
and the normal cone is
\begin{equation}
N_{[0,2]}(x)=
\begin{cases}
(-\infty,0], & x=0,\\
\{0\}, & 0<x<2,\\
[0,\infty), & x=2.
\end{cases}
\label{eq:normal}
\end{equation}
Introduce multipliers $\lambda_0,\lambda_2\ge0$ and write
\begin{equation}
\Xi\dot x=-\frac{\partial G}{\partial x}+\lambda_0-\lambda_2,
\quad
C\dot q=-\frac{\partial G}{\partial q},
\label{eq:complement}
\end{equation}
with
\begin{equation}
\lambda_0x=0,
\quad
\lambda_2(2-x)=0.
\label{eq:complement2}
\end{equation}
At an interior point $0<x<2$, $\lambda_0=\lambda_2=0$, and the linear system \eqref{eq:xdot} and \eqref{eq:qdot} is recovered. At $x=0$, if the free force points outward, $\lambda_0$ cancels the outward component and $\dot x=0$. At $x=2$, if the free force points outward, $\lambda_2$ cancels the outward component and $\dot x=0$. The projected system retains the Lyapunov inequality
\begin{equation}
\frac{\mathrm{d}G}{\mathrm{d}t}=-\Xi\dot x^2-C\dot q^2\le0.
\label{eq:energyproj}
\end{equation}
so the boundary multipliers do not inject energy into the physical interval. The main target $x_f=1.9$ lies in the interior. If $A>0$ and $x_\infty>x_f$, the trajectory satisfies $0<x(t)<1.9<2$ before the first hit of $x_f$. The boundary multipliers are zero in that time interval. The double-exponential interior solution therefore gives the exact first-passage time for the projected dynamics. If $x_\infty>2$, the unconstrained curve is plotted only up to the first hit of $x=2$. The later evolution is a projected dynamics with $x=2$ and internal relaxation of $q$. The unconstrained value $x_\infty>2$ does not mean that the spherical cap exceeds the full sphere.

\section{Long-time thresholds and phase boundaries}
\label{sec:thresholds}

\begin{figure}[t]
\centering
\includegraphics[width=\linewidth]{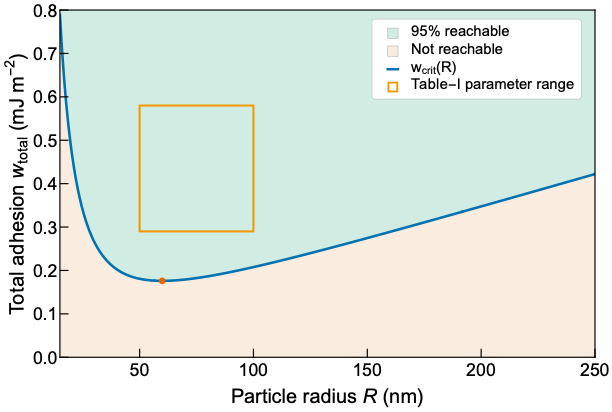}
\caption{Reachability boundary in the $(R,w_{\mathrm{total}})$ plane for 95\% wrapping. The blue curve is $w_{\mathrm{crit}}(R)$; the green region above it is reachable in the long-time limit, and the orange region below it is not. The orange rectangle marks the Table-I parameter range, and the minimum of the boundary is at $R_*\approx59.9\,\mathrm{nm}$.}
\label{fig:figure11}
\end{figure}

\subsection{Total adhesion threshold}
\label{subsec:wcrit}

The long-time limit is
\begin{equation}
x_\infty=\frac{2\pi R^2w_{\mathrm{total}}-4\pi\kappa}{2\pi\gamma R^2+\chi E_\infty R^3}.
\end{equation}
The target $x_f$ is reachable in finite time when $x_\infty>x_f$. The critical condition $x_\infty=x_f$ gives
\begin{equation}
2\pi R^2w_{\mathrm{total}}-4\pi\kappa
=x_f\left(2\pi\gamma R^2+\chi E_\infty R^3\right).
\end{equation}
Dividing by $2\pi R^2$ gives
\begin{equation}
w_{\mathrm{crit}}(R)
=\frac{2\kappa}{R^2}+\gamma x_f+\frac{\chi E_\infty R x_f}{2\pi}.
\label{eq:wcrit}
\end{equation}
The first term is the bending cost, the second is the tension cost, and the third is the long-time cortical elastic cost. The competition between the decreasing bending term and the increasing cortical term gives a finite optimal radius. The complete threshold algebra is collected in Appendix~\hyperref[app:threshold]{E}, and the reachability boundary in the $(R,w_{\mathrm{total}})$ plane is shown in Fig.~\ref{fig:figure11}.

\begin{figure}[t]
\centering
\includegraphics[width=\linewidth]{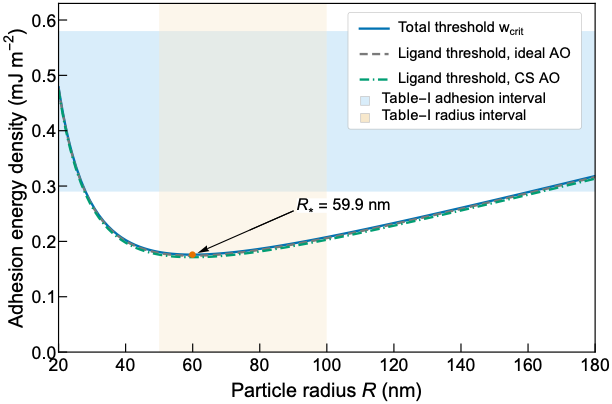}
\caption{Total adhesion threshold for 95\% wrapping and the minimum ligand adhesion at $r=15\,\mathrm{nm}$. The solid curve is the required total $w_{\mathrm{crit}}(R)$; the dashed and dash-dotted curves are the ligand thresholds obtained by subtracting the ideal and CS pressure-rescaled depletion contributions, respectively. All curves are U-shaped and share the minimum at $R_*\approx59.9\,\mathrm{nm}$; the shallow bands mark the Table-I radius and adhesion intervals.}
\label{fig:figure10}
\end{figure}

The required ligand adhesion is
\begin{equation}
w_{\mathrm{LR,min}}
=\max\left\{0,
\frac{2\kappa}{R^2}+\gamma x_f+\frac{\chi E_\infty R x_f}{2\pi}
-w_{\mathrm{dep}}
\right\}.
\label{eq:wLRmin}
\end{equation}
For fixed $r$, $c$, and $T$, the depletion density $w_{\mathrm{dep}}$ is independent of $R$. The curves for total threshold and ligand threshold are therefore vertical shifts of one another; the threshold-minimizing radius is unchanged by this shift. Figure~\ref{fig:figure10} compares the required total adhesion with the minimum ligand adhesion obtained by subtracting the ideal or CS pressure-rescaled depletion contribution.

\subsection{Threshold-minimizing radius and minimum threshold}
\label{subsec:Rstar}

Differentiating \eqref{eq:wcrit} with respect to $R$ gives
\begin{equation}
\frac{\mathrm{d}w_{\mathrm{crit}}}{\mathrm{d}R}
=-\frac{4\kappa}{R^3}+\frac{\chi E_\infty x_f}{2\pi}.
\label{eq:dwcrit}
\end{equation}
Setting this to zero gives the unique positive stationary point
\begin{equation}
R_*=\left(\frac{8\pi\kappa}{\chi E_\infty x_f}\right)^{1/3}.
\label{eq:Rstar}
\end{equation}
The second derivative
\begin{equation}
\frac{\mathrm{d}^2w_{\mathrm{crit}}}{\mathrm{d}R^2}
=\frac{12\kappa}{R^4}>0
\label{eq:d2wcrit}
\end{equation}
shows that $R_*$ is the unique global minimum on the positive radius axis. At $R_*$,
\begin{equation}
\frac{2\kappa}{R_*^2}
=\frac{\chi E_\infty x_f R_*}{4\pi},
\label{eq:bending-at-Rstar}
\end{equation}
so the minimum total threshold is
\begin{equation}
w_{\mathrm{crit,min}}
=\gamma x_f+\frac{3\chi E_\infty x_f R_*}{4\pi}
=\gamma x_f+\frac{6\kappa}{R_*^2}.
\label{eq:wcritmin}
\end{equation}
At the optimal radius, the cortical linear term is twice the bending term, and the sum of the two is three times the bending term.

\begin{figure}[t]
\centering
\includegraphics[width=\linewidth]{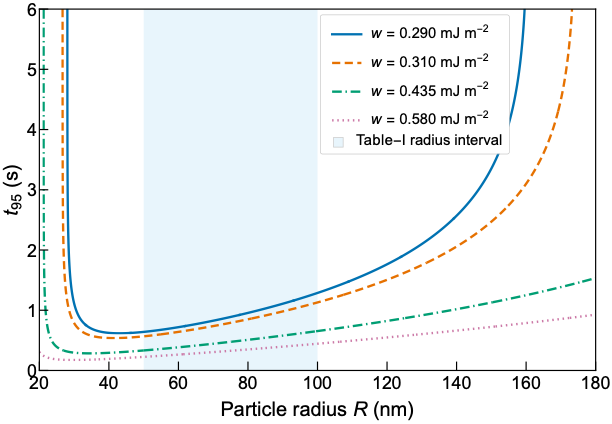}
\caption{Exact 95\% first-passage time versus particle radius for total adhesion densities $w_{\mathrm{total}}=0.290$, $0.310$, $0.435$, and $0.580\,\mathrm{mJ\,m^{-2}}$. Each curve is U-shaped; the minima are at $R\approx42.96$, $41.42$, $34.42$, and $29.61\,\mathrm{nm}$ with $t_{95}\approx0.617$, $0.537$, $0.284$, and $0.174\,\mathrm{s}$, respectively. The light band marks the Table-I radius interval; gaps and endpoints indicate that the long-time limit does not reach $x_f=1.9$.}
\label{fig:figure08}
\end{figure}

For the representative parameters $\kappa=8.18\times10^{-20}\,\mathrm{J}$, $E_\infty=1999.9600008\,\mathrm{Pa}$, $x_f=1.9$, and $\chi=16\sqrt{2}/9$,
\begin{equation}
R_*=59.88\,\mathrm{nm},
\qquad
w_{\mathrm{crit,min}}=0.176005\,\mathrm{mJ\,m^{-2}}.
\end{equation}
With $r=15\,\mathrm{nm}$, the minimum ligand threshold is $0.174141\,\mathrm{mJ\,m^{-2}}$ for ideal pressure and $0.171252\,\mathrm{mJ\,m^{-2}}$ for CS pressure rescaling.

\subsection{Fixed adhesion and two positive radius roots}
\label{subsec:cubic}

\begin{figure}[t]
\centering
\includegraphics[width=\linewidth]{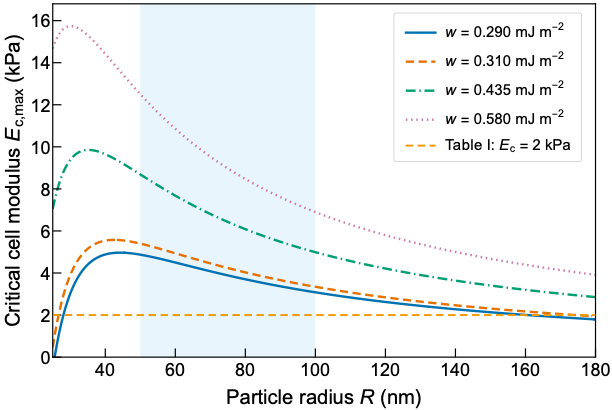}
\caption{Maximum cell modulus compatible with 95\% wrapping for total adhesion densities $w_{\mathrm{total}}=0.290$, $0.310$, $0.435$, and $0.580\,\mathrm{mJ\,m^{-2}}$. The region below each curve is reachable in the long-time limit. The curves peak at finite radii (about $44.3$, $42.6$, $35.2$, and $30.2\,\mathrm{nm}$); stronger adhesion allows a larger tolerable modulus and shifts the peak to smaller $R$. The horizontal dashed line marks the Table-I value $E_c=2\,\mathrm{kPa}$. The plotted boundary is the long-time condition $x_\infty=x_f$; the first-passage time diverges as this boundary is approached from the reachable side.}
\label{fig:figure12}
\end{figure}

For fixed total adhesion $w_{\mathrm{total}}$, define
\begin{equation}
W=w_{\mathrm{total}}-\gamma x_f,
\quad
C_R=\frac{\chi E_\infty x_f}{2\pi}>0.
\label{eq:WCR}
\end{equation}
The boundary equation $W=2\kappa/R^2+C_RR$ becomes, after multiplying by $R^2$,
\begin{equation}
f(R)=C_RR^3-WR^2+2\kappa=0.
\label{eq:cubic}
\end{equation}
If $W\le0$, the left-hand side is strictly positive for $R>0$, so there is no reachable window. For $W>0$, the positive stationary point of $f$ is
\begin{equation}
R_m=\frac{2W}{3C_R},
\label{eq:Rm}
\end{equation}
where $f''(R_m)=2W>0$, so $R_m$ is the unique minimum on the positive axis. Evaluating $f$ at $R_m$ gives
\begin{equation}
f(R_m)=2\kappa-\frac{4W^3}{27C_R^2}.
\label{eq:fRm}
\end{equation}
Because $f(0)=2\kappa>0$ and $f(R)\to+\infty$, the cubic has two positive roots when
\begin{equation}
W^3>\frac{27}{2}\kappa C_R^2.
\label{eq:tworoots}
\end{equation}
At equality the two positive roots merge at $R_{\mathrm{fold}}=2W/(3C_R)$. The two roots define the finite size window $R_{\min}<R<R_{\max}$ for fixed adhesion. The boundary in the $(R,w_{\mathrm{total}})$ plane is a single U-shaped curve $w_{\mathrm{total}}=w_{\mathrm{crit}}(R)$; the two roots appear when a horizontal line $w_{\mathrm{total}}=\text{constant}$ is drawn through this curve. The radius dependence of the exact 95\% first-passage time is shown in Fig.~\ref{fig:figure08}.

\begin{figure}[t]
\centering
\includegraphics[width=\linewidth]{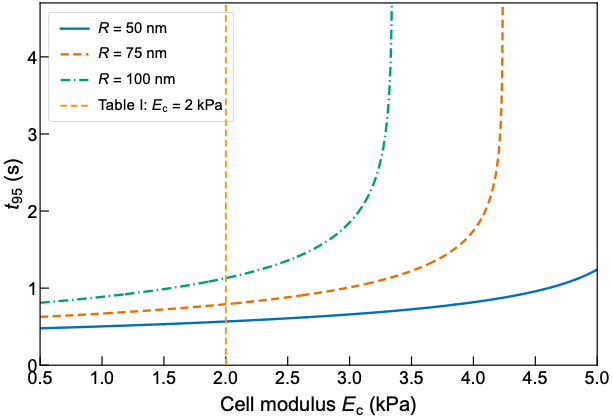}
\caption{Exact 95\% first-passage time versus cell modulus for $R=50$, $75$, and $100\,\mathrm{nm}$ at total adhesion $w_{\mathrm{total}}=3.1\times10^{-4}\,\mathrm{J\,m^{-2}}$. The times increase with $E_c$ and diverge near the long-time reachability limits; for $R=75$ and $100\,\mathrm{nm}$ the curves terminate at about $4.234$ and $3.339\,\mathrm{kPa}$, respectively, whereas $R=50\,\mathrm{nm}$ remains reachable up to $5\,\mathrm{kPa}$. The vertical dashed line marks the Table-I value $E_c=2\,\mathrm{kPa}$. The last plotted samples are finite-time sampling points; at the actual boundary $x_\infty=x_f$, $t_{95}$ diverges.}
\label{fig:figure09}
\end{figure}

\subsection{Critical cell modulus}
\label{subsec:Ecmax}

The long-time threshold can be solved for the maximum allowed long-time modulus:
\begin{equation}
E_{\infty,\max}
=\frac{2\pi}{\chi x_fR}
\left(
w_{\mathrm{total}}-\frac{2\kappa}{R^2}-\gamma x_f
\right).
\label{eq:Einfmax}
\end{equation}
The bracket must be positive. Since
\begin{equation}
E_\infty=\frac{E_vE_c}{E_v+E_c},
\end{equation}
solving for $E_c$ gives
\begin{equation}
E_{c,\max}
=\frac{E_vE_{\infty,\max}}{E_v-E_{\infty,\max}},
\label{eq:Ecmax}
\end{equation}
with $0<E_{\infty,\max}<E_v$. For $E_{\infty,\max}\ge E_v$, all finite cell moduli satisfy the long-time bound and the critical cell modulus is unbounded in the reduced model. The maximum compatible cell modulus is shown in Fig.~\ref{fig:figure12}, and the exact $95\%$ first-passage time versus $E_c$ is shown in Fig.~\ref{fig:figure09}.

For fixed $w_{\mathrm{total}}$, write
\begin{equation}
E_{\infty,\max}(R)
=\frac{2\pi}{\chi x_f}
\left[
\frac{w_{\mathrm{total}}-\gamma x_f}{R}
-\frac{2\kappa}{R^3}
\right].
\label{eq:EinfmaxR}
\end{equation}
Setting the derivative with respect to $R$ to zero gives
\begin{equation}
R_{E,\max}=\sqrt{\frac{6\kappa}{w_{\mathrm{total}}-\gamma x_f}},
\label{eq:REmax}
\end{equation}
provided $w_{\mathrm{total}}>\gamma x_f$. This radius is distinct from the threshold-minimizing radius $R_*$. The radius $R_*$ minimizes the required adhesion at fixed modulus; the radius $R_{E,\max}$ maximizes the tolerable modulus at fixed adhesion; the time-optimal radius is a third quantity, discussed below.

\subsection{Time-optimal radius}
\label{subsec:time-optimal}

The first-passage time is controlled by $A(R)$, $\Xi(R)$, $C(R)$, $K_\gamma(R)$, $K_\infty(R)$, $K_1(R)$, and the two eigenvalues and velocity residues. The static threshold-minimizing radius $R_*$ does not minimize the first-passage time. For $w_{\mathrm{total}}=0.310\,\mathrm{mJ\,m^{-2}}$, the time-optimal radius is about $41.42\,\mathrm{nm}$, whereas the static threshold-minimizing radius is $59.88\,\mathrm{nm}$. The difference arises from the competition among the driving coefficient $A(R)$, the external drag $\Xi(R)$, the internal dissipation $C(R)$, and the stiffnesses. For the representative parameters, the wrapping time is predominantly controlled by viscoelastic relaxation and is only weakly sensitive to the external Stokes drag.

\begin{figure*}[t]
\centering
\includegraphics[width=\linewidth]{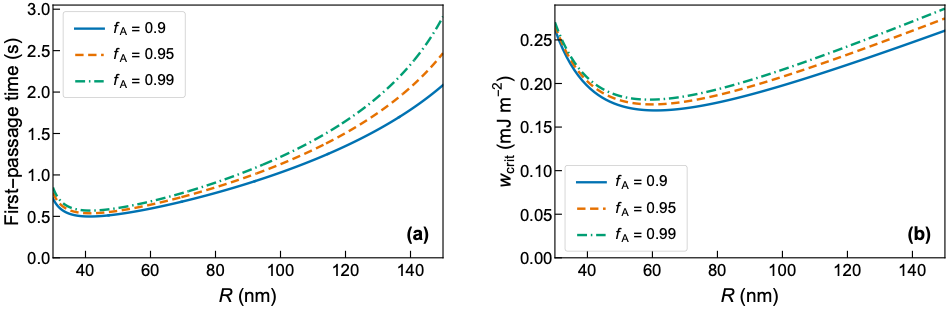}
\caption{Sensitivity to the spherical-area target $f_A=0.90$, $0.95$, and $0.99$. (a) Exact first-passage time versus $R$; in the common reachable range the ordering $t_{90}<t_{95}<t_{99}$ is preserved, and the time minima are near $41.3$--$41.5\,\mathrm{nm}$. (b) Static total adhesion threshold versus $R$; the threshold increases with $f_A$, while the threshold-minimizing radius shifts only slowly from about $61.09$ to $59.14\,\mathrm{nm}$.}
\label{fig:figure15}
\end{figure*}

\section{The 95\% spherical-area criterion and its sensitivity}
\label{sec:criterion95}

\begin{table}[t]
\centering
\caption{Sensitivity to the spherical-area target. The time-optimal radius differs from the static threshold-minimizing radius.}
\label{tab:sensitivity}
\begin{tabular}{cccc}
\toprule
$f_A$ & $t_{\min}$ ($\mathrm{s}$) & $R_{t,\min}$ ($\mathrm{nm}$) & $w_{\mathrm{crit,min}}$ ($\mathrm{mJ\,m^{-2}}$)\\
\midrule
0.90 & 0.49843 & 41.31 & 0.169096\\
0.95 & 0.53729 & 41.42 & 0.176005\\
0.99 & 0.56973 & 41.51 & 0.181471\\
\bottomrule
\end{tabular}
\end{table}

The wrapped area fraction is $f_A=x/2$. The 95\% criterion is $f_A=0.95$, $x_f=1.9$. For a general target fraction $f\in(0,1)$, $x_f=2f$, and the total adhesion threshold becomes
\begin{equation}
w_{\mathrm{crit}}(R;f)
=\frac{2\kappa}{R^2}+2f\gamma+\frac{\chi f}{\pi}E_\infty R.
\label{eq:wcritf}
\end{equation}
Differentiating with respect to $f$ gives
\begin{equation}
\frac{\partial w_{\mathrm{crit}}}{\partial f}
=2\gamma+\frac{\chi E_\infty R}{\pi}>0.
\label{eq:dwcritf}
\end{equation}
The required adhesion increases with the target fraction. The threshold-minimizing radius satisfies
\begin{equation}
R_*(f)=\left(\frac{4\pi\kappa}{\chi fE_\infty}\right)^{1/3},
\label{eq:Rstarf}
\end{equation}
so $R_*(f)\propto f^{-1/3}$. The radius changes slowly with $f$, while the threshold itself increases nearly linearly with $f$.

For fixed parameters and $f_1<f_2$, the target coordinates satisfy $x_{f_1}<x_{f_2}$. Since $x(t)$ is strictly increasing, the first-passage times satisfy $t_{f_1}<t_{f_2}$. In the representative scan, the shortest first-passage times are approximately $0.49843\,\mathrm{s}$, $0.53729\,\mathrm{s}$, and $0.56973\,\mathrm{s}$ for $f_A=0.90$, $0.95$, and $0.99$. The corresponding static minimum thresholds are $0.169096$, $0.176005$, and $0.181471\,\mathrm{mJ\,m^{-2}}$. The corresponding threshold-minimizing radii are about $61.09$, $59.88$, and $59.14\,\mathrm{nm}$. These values are collected in Table~\ref{tab:sensitivity}, and the corresponding sensitivity curves are shown in Fig.~\ref{fig:figure15}.

The 95\% criterion is an operational near-complete threshold. Midya, Auth and Gompper use $f_w=0.95$ as a stable deep wrapping configuration while keeping the complete-wrapped state as a separate state in their phase diagram \cite{midya2023}. The present model uses $x_f=1.9$ for the main calculation and reports $x_f=1.8$ and $x_f=1.98$ for sensitivity. The model does not describe neck formation, membrane fusion, dynamin scission, or intracellular release. The time $t_{95}$ is a first-passage time to a prescribed spherical area fraction, not a topological completion time.

In polymorphic models of receptor-mediated endocytosis, a common operational conventions to treat particles with a wrapping fraction in the literature vary. A 90\% operational convention has been adopted in continuous models \cite{Chaudhuri2011}, and a 95\% operational convention has been used in coarse-grained studies \cite{Paesani2026}. To be conservative, we adopt the 95\% criterion as our operational completion threshold and compare 90\%, 95\%, and 99\% as a sensitivity check. This convention reflects the physical picture that wrapping is essentially complete before neck scission, but it is not a statement that 95\% coincides with a measured invagination depth or with topological completion.

\section{Representative numerical values}
\label{sec:numerics}

The Table-I parameters used in this manuscript are collected in Table~\ref{tab:params}. The derived single-mode standard linear solid parameters are collected in Table~\ref{tab:sls}. The 95\% sensitivity values are collected in Table~\ref{tab:sensitivity}. The parameter-grid values are collected in Table~\ref{tab:parametergrid}.

The 95\% first-passage time at the representative point is
\begin{equation}
t_{95}=0.5668368839\,\mathrm{s}.
\end{equation}
The unconstrained long-time limit is $x_\infty=4.0281152277$. The fast eigenvalue is $\lambda_-=-1.3338194590\times10^{10}\,\mathrm{s^{-1}}$, and the slow eigenvalue is $\lambda_+=-1.1255998593\,\mathrm{s^{-1}}$. The fast time is about $7.50\times10^{-11}\,\mathrm{s}$, and the slow time is about $0.8884\,\mathrm{s}$. The large separation of time scales follows from the large ratio $E_v/E_c=5\times10^4$ and from the coupling of the stiff instantaneous branch to the external drag.

The radius scan for $w_{\mathrm{total}}=0.310\,\mathrm{mJ\,m^{-2}}$ gives a reachable interval approximately $26.76$ to $174.30\,\mathrm{nm}$. The shortest 95\% time is about $0.5373\,\mathrm{s}$ at $R\approx41.42\,\mathrm{nm}$. The static threshold-minimizing radius at the same parameters is $R_*=59.88\,\mathrm{nm}$. The time-optimal radius and the static threshold-minimizing radius are distinct. The radius dependence is shown in Fig.~\ref{fig:figure08}.

\begin{table*}[t]
\centering
\caption{Fixed-ligand-chemistry Table-I parameter grid. The ligand adhesion is calibrated so that the ideal AO reference at $r=15\,\mathrm{nm}$ gives total adhesion $w_{\mathrm{total}}=3.1\times10^{-4}\,\mathrm{J\,m^{-2}}$. The CS depletion contribution is then added at each $r$. This table is not the fixed-total-adhesion representative point used in the text.}
\label{tab:parametergrid}
\begin{tabular}{lllllll}
\toprule
$R$ (nm) & $r$ (nm) & $\phi$ & $w_{\mathrm{dep}}$ (mJ m$^{-2}$) & $w_{\mathrm{total}}$ (mJ m$^{-2}$) & $x_\infty$ & $t_{95}$ (s)\\
\midrule
50 & 5 & 0.007854 & 0.000641 & 0.308777 & 4.00798 & 0.57083\\
75 & 5 & 0.007854 & 0.000641 & 0.308777 & 3.46498 & 0.79665\\
100 & 5 & 0.007854 & 0.000641 & 0.308777 & 2.90308 & 1.13810\\
50 & 10 & 0.062832 & 0.001610 & 0.309746 & 4.02394 & 0.56766\\
75 & 10 & 0.062832 & 0.001610 & 0.309746 & 3.47698 & 0.79246\\
100 & 10 & 0.062832 & 0.001610 & 0.309746 & 2.91270 & 1.13143\\
50 & 15 & 0.212058 & 0.004753 & 0.312889 & 4.07570 & 0.55762\\
75 & 15 & 0.212058 & 0.004753 & 0.312889 & 3.51592 & 0.77918\\
100 & 15 & 0.212058 & 0.004753 & 0.312889 & 2.94390 & 1.11034\\
50 & 20 & 0.502655 & 0.032894 & 0.341030 & 4.53921 & 0.48174\\
75 & 20 & 0.502655 & 0.032894 & 0.341030 & 3.86455 & 0.67813\\
100 & 20 & 0.502655 & 0.032894 & 0.341030 & 3.22329 & 0.95345\\
\bottomrule
\end{tabular}
\end{table*}

The cell-modulus scan at $w_{\mathrm{total}}=0.310\,\mathrm{mJ\,m^{-2}}$ gives finite positive 95\% times for $R=50\,\mathrm{nm}$ up to $E_c=5\,\mathrm{kPa}$. For $R=75\,\mathrm{nm}$, the curve ends at about $E_c=4.234\,\mathrm{kPa}$, with $t_{95}\approx4.031\,\mathrm{s}$ at the last reachable sample. For $R=100\,\mathrm{nm}$, the curve ends at about $E_c=3.339\,\mathrm{kPa}$, with $t_{95}\approx4.514\,\mathrm{s}$ at the last reachable sample. These last plotted samples are finite-time sampling points, not the critical points; the termination is set by $x_\infty=x_f$, and the first-passage time diverges as the boundary is approached. The corresponding cell-modulus dependence is shown in Fig.~\ref{fig:figure09}.

\section{Discussions}
\label{sec:discussion}

The reduced description above isolates the minimal ingredients needed for depletion-assisted capture and viscoelastic wrapping. Its conclusions are therefore conditional on the following model assumptions, which also indicate the natural extensions. The AO geometry in \eqref{eq:VAO} is exact for ideal hard-sphere crowders before contact, and $U_{\mathrm{AO}}=-\Pi V_{\mathrm{AO}}$ is exact only within that model. Real crowders have shape, polydispersity, flexibility, charge, and specific interactions. The CS replacement changes the bulk pressure while retaining the ideal overlap geometry. By construction, it does not aim to reproduce finite-density correlations, layering, bridging, or curvature-dependent direct correlation functions. At fixed $c=1.5\times10^{22}\,\mathrm{m^{-3}}$, $r=20\,\mathrm{nm}$ gives $\phi=0.502655$, above the hard-sphere fluid freezing fraction $\phi_f\simeq0.494$. The high-radius endpoint is therefore retained only as an extrapolation of the fluid branch.

The capture stage uses the exact sphere-plane overlap volume, whereas the wrapping stage uses the local parallel-plate density $w_{\mathrm{dep}}=2\Pi r$. These are different reductions: $U_{\mathrm{AO}}(0)$ is not divided by an arbitrary area to define $w_{\mathrm{dep}}$, and the two stages are not coupled through a single reaction coordinate. The deterministic times \eqref{eq:tbulk} and \eqref{eq:tB} are conditional on entry into the depletion layer and exclude stochastic first entry, receptor binding waiting times, and a full stochastic total endocytosis time. The Brenner correction assumes a rigid sphere near a rigid no-slip plane; a deformable membrane with in-plane viscosity, slip, or solvent permeation would require boundary-integral or lubrication treatment. The bulk Stokes and Brenner results are reference estimates.

The wrapping geometry uses one spherical-cap coordinate and assumes a circular contact line, axial symmetry, a spherical particle, and membrane attachment in the contact region. It does not resolve wrinkles, non-axisymmetric instabilities, eccentricity, nonuniform in-plane tension, or local detachment; for nonspherical particles a single area fraction is insufficient. The tension energy $G_\gamma=\pi\gamma R^2x^2$ assumes a constant tension reservoir and a flat outer membrane, and $G_\kappa=4\pi\kappa x$ assumes zero spontaneous curvature with the Gaussian boundary contribution absorbed into the reduced description. Coat proteins, lipid asymmetry, or curvature-generating proteins would add spontaneous-curvature terms, and the model does not predict scission barriers. The adhesion density $w_{\mathrm{total}}$ is uniform and constant; real receptors diffuse, cluster, deplete, and bind with force-dependent kinetics, and the depletion contribution can change with local gap and crowder redistribution~\cite{wu2026multivalent}.

The single-mode standard linear solid contains one instantaneous modulus, one long-time modulus, and one material time scale. It captures the internal-variable memory needed to represent the creep compliance of \cite{Liu2026e2} and provides a transparent reference description. A multimode extension would replace $q$ by $q_i$ with
\begin{equation}
\begin{aligned}
G_{\mathrm{ve}}&=\frac{1}{2}K_\infty x^2+\sum_i\frac{1}{2}K_i(x-q_i)^2,
\\
\Phi_{\mathrm{int}}&=\sum_i\frac{1}{2}C_i\dot q_i^2,
\end{aligned}
\end{equation}
which remains a linear gradient flow but yields a multiexponential first-passage trajectory. The dissipation matrix is diagonal; cross coefficients allowed by linear irreversible thermodynamics would preserve the energy inequality but require separate analysis of eigenvalues, positive-system structure, and residue signs. The drag $\Xi=6\pi\mu R^3$ is a single-mode effective coefficient, not a full membrane-fluid boundary-value result. The first-passage time depends on both the external drag and the internal dissipation. For the representative parameters, it is predominantly controlled by viscoelastic relaxation and is only weakly sensitive to the external Stokes drag.

Table~\ref{tab:parametergrid} uses a fixed-ligand-chemistry convention: $w_{\mathrm{LR}}$ is calibrated so that the ideal-AO reference at $r=15\,\mathrm{nm}$ gives $w_{\mathrm{total}}=3.1\times10^{-4}\,\mathrm{J\,m^{-2}}$, and the CS depletion contribution is then added at each $r$. Thus the $r=15\,\mathrm{nm}$ CS entry has $w_{\mathrm{total}}=0.312889\,\mathrm{mJ\,m^{-2}}$ and $t_{95}=0.55762\,\mathrm{s}$, slightly different from the representative fixed-total-adhesion point $w_{\mathrm{total}}=0.310\,\mathrm{mJ\,m^{-2}}$ and $t_{95}=0.56684\,\mathrm{s}$. The two are not inconsistent; they answer different adhesion-accounting questions.

The model is deterministic and describes mean RO dynamics. A fluctuation-dissipation-consistent stochastic extension~\cite{wu2026single} adds noise,
\begin{equation}
\Gamma\dot{\boldsymbol\eta}=-\nabla G+\sqrt{2k_{\mathrm{B}}T\Gamma}\boldsymbol\xi(t),
\end{equation}
making the first-passage time a distribution; the present $t_{95}$ does not include this variance or rare events near the reachability boundary. Active processes such as actin polymerization, motor proteins, active tension regulation, and internalization machinery inject energy and can violate pure free-energy decay. The conclusions therefore apply to passive dynamics or to conditions in which active effects are absorbed into effective constant adhesion, and the model does not establish that real cell trajectories are always monotone or nonoscillatory.

The unconstrained limit $x_\infty>2$ means that the free force still points outward at the full-sphere boundary; the physical system must activate a normal reaction, a neck degree of freedom, or a topological change at $x=2$, and $x_\infty=4$ is not a prediction of $200\%$ wrapping. The threshold \eqref{eq:wcrit} is conditional on fixed $\kappa$, $\gamma$, $E_\infty$, and $f_A$ and excludes nucleation barriers, contact-line pinning, discrete receptors, and active processes: reachability means only that the unconstrained long-time limit exceeds the target, not that every microscopic initial state reaches it without a barrier.

Additional adhesive structures and active cortical flows can contribute to the initial approach and the wrapped state \cite{GuptaSantangeloPattesonSchwarz2025,Song2025,Huang2013,Nividha2026enzyme,wu2018getting,Jacobson2019lateral}. Frameworks developed for amoeboid swimming in confined geometries \cite{Wu2015,Wu2016,Farutin2019}, molecular-paddling amoeboid swimming in lymphocytes \cite{Aoun2020}, and chemotaxis-guided cell swimming \cite{Wang2021} provide natural starting points for describing such active flows. These effects can significantly modify engulfment, and the reduced passive model developed here provides a transparent basis for extending the present description to active cortical flow effects.

\section{Conclusion}
\label{sec:conclusion}

A two-stage description of receptor-mediated endocytosis has been developed. The capture stage uses the exact ideal hard-sphere AO sphere-plane overlap volume, the depletion potential and force, CS pressure rescaling, bulk Stokes approach time, and Brenner normal wall drag. The wrapping stage uses a spherical-cap coordinate, Helfrich bending, membrane tension, local adhesion, and a single-mode standard linear solid with one internal strain variable. The creep compliance of Liu et~al. \cite{Liu2026e2} is mapped exactly to a long-time spring in parallel with a Maxwell branch. The mapping preserves the creep retardation time and the stress relaxation time.

The wrapping dynamics is a two-variable RO gradient flow with diagonal dissipation. The linear system has two real negative eigenvalues and an explicit double-exponential solution. Two independent arguments give strict monotonicity of the wrapping coordinate for positive initial drive. The first-passage equation has a unique finite root when the target lies below the unconstrained long-time limit. The projected dynamics on \(0\le x\le2\) is specified by normal-cone conditions. The 95\% spherical-area criterion is treated as an operational near-complete threshold, and the 90\%, 95\%, and 99\% cases are compared.

Long-time thresholds, the threshold-minimizing radius, the critical cell modulus, the size window, and the phase boundaries follow from the same free energy. For the Table-I representative parameters, the 95\% first-passage time at \(R=50\,\mathrm{nm}\) and total adhesion \(3.1\times10^{-4}\,\mathrm{J\,m^{-2}}\) is about \(0.567\,\mathrm{s}\). The static threshold-minimizing radius is \(59.88\,\mathrm{nm}\), while the time-optimal radius at \(w_{\mathrm{total}}=0.310\,\mathrm{mJ\,m^{-2}}\) is about \(41.42\,\mathrm{nm}\). The capture time is in the sub-millisecond range in the deterministic estimate for the chosen entry offset $\epsilon=0.01\,\mathrm{nm}$. The staged treatment---with stochastic capture and multivalent stabilization treated separately in Refs.~\cite{wu2026single,wu2026multivalent}---the single-mode viscoelastic spectrum, the diagonal dissipation, the pressure-only crowding correction, and the reduced spherical-cap geometry define the scope of the conclusions. 

\begin{acknowledgments}
Z.O. is supported by the Major Program of National Natural Science Foundation of China under Grant No. 22193032. H.W. is supported by the General Program of NSFC under Grant No. 12374210, the open research fund of Songshan Lake Materials Laboratory No. 2023SLABFN20, and the startup fund No. WIUCASQD2022005 from the Wenzhou Institute, University of Chinese Academy of Sciences.
\end{acknowledgments}

\appendix
\renewcommand{\thesection}{\Alph{section}}
\setcounter{section}{0}
\renewcommand{\theequation}{S\arabic{equation}}
\setcounter{equation}{0}
\renewcommand{\thefigure}{S\arabic{figure}}
\setcounter{figure}{0}

\section{Appendix A: AO overlap volume}
\label{app:AO}

The cap height is \(q=2r-d\). The cap volume of a sphere of radius \(A=R+r\) and height \(q\) is
\begin{equation}
V_{\mathrm{cap}}
=\int_0^q\pi(2As-s^2)\,\mathrm{d}s
=\pi q^2\left(A-\frac{q}{3}\right).
\end{equation}
Substituting \(A=R+r\) and \(q=2r-d\),
\begin{equation}
V_{\mathrm{AO}}(d)=\frac{\pi}{3}(2r-d)^2(3R+r+d).
\end{equation}
The derivative is
\begin{equation}
V_{\mathrm{AO}}'(d)=-\pi(2r-d)(2R+d).
\end{equation}
The force along increasing gap is
\begin{equation}
F_d=-\frac{\mathrm{d}U_{\mathrm{AO}}}{\mathrm{d}d}
=-\pi\Pi(2r-d)(2R+d).
\end{equation}
The magnitude is
\begin{equation}
|F_{\mathrm{AO}}|=\pi\Pi(2r-d)(2R+d).
\end{equation}

\section{Appendix B: Spherical-cap geometry}
\label{app:cap}

The spherical area element is
\begin{equation}
\mathrm{d}A=R^2\sin\theta\,\mathrm{d}\theta\,\mathrm{d}\phi.
\end{equation}
Integrating from the pole to the contact angle \(\theta\),
\begin{equation}
A_c=\int_0^{2\pi}\int_0^\theta R^2\sin\theta\,\mathrm{d}\theta\,\mathrm{d}\phi
=2\pi R^2(1-\cos\theta).
\end{equation}
With \(x=1-\cos\theta\),
\begin{equation}
A_c=2\pi R^2x,
\qquad
f_A=\frac{A_c}{4\pi R^2}=\frac{x}{2}.
\end{equation}
The contact-line radius is
\begin{equation}
\rho=R\sin\theta.
\end{equation}
Using \(\cos\theta=1-x\),
\begin{equation}
\sin^2\theta=2x-x^2.
\end{equation}
Thus
\begin{equation}
\rho=R\sqrt{2x-x^2}.
\end{equation}
The projected area is
\begin{equation}
A_p=\pi\rho^2=\pi R^2(2x-x^2).
\end{equation}
The excess area is
\begin{equation}
\Delta A=A_c-A_p=\pi R^2x^2.
\end{equation}
The Helfrich bending energy for a sphere of radius \(R\) is
\begin{equation}
G_\kappa=\frac{\kappa}{2}\left(\frac{2}{R}\right)^2A_c
=4\pi\kappa x.
\end{equation}
The tension energy is
\begin{equation}
G_\gamma=\gamma\Delta A=\pi\gamma R^2x^2.
\end{equation}
The adhesion energy is
\begin{equation}
G_w=-w_{\mathrm{total}}A_c=-2\pi R^2w_{\mathrm{total}}x.
\end{equation}

\section{Appendix C: Standard linear solid mapping}
\label{app:sls}

The creep compliance is
\begin{equation}
J(t)=\frac{1}{E_v}+\frac{1}{E_c}\left(1-e^{-t/\tau_c}\right),
\quad
\tau_c=\frac{\eta_c}{E_c}.
\end{equation}
The internal-variable model has
\begin{equation}
\psi=\frac{1}{2}E_\infty x^2+\frac{1}{2}E_1(x-q)^2,
\quad
\phi_q=\frac{1}{2}\eta_M\dot q^2.
\end{equation}
The stress is
\begin{equation}
\sigma=E_\infty x+E_1(x-q).
\end{equation}
Stationarity gives
\begin{equation}
\eta_M\dot q=E_1(x-q).
\end{equation}
The constitutive equation is
\begin{equation}
\sigma+\tau_{\mathrm{rel}}\dot\sigma
=E_\infty x+E_0\tau_{\mathrm{rel}}\dot x,
\quad
\tau_{\mathrm{rel}}=\frac{\eta_M}{E_1}.
\end{equation}
For a step stress, the creep compliance is
\begin{equation}
J_{\mathrm{SLS}}(t)=\frac{1}{E_\infty}
-\left(\frac{1}{E_\infty}-\frac{1}{E_0}\right)e^{-t/\tau_c},
\end{equation}
with
\begin{equation}
\tau_c=\frac{\eta_M E_0}{E_1E_\infty}.
\end{equation}
Matching gives
\begin{equation}
E_0=E_v,
\quad
E_\infty=\frac{E_vE_c}{E_v+E_c},
\quad
E_1=E_0-E_\infty,
\end{equation}
\begin{equation}
\eta_M=\tau_c\frac{E_1E_\infty}{E_0}
=\eta_c\frac{E_v^2}{(E_v+E_c)^2},
\end{equation}
\begin{equation}
\tau_{\mathrm{rel}}=\frac{\eta_M}{E_1}
=\frac{\eta_c}{E_v+E_c}.
\end{equation}

\section{Appendix D: Matrix spectrum and residues}
\label{app:matrix}

The matrix is
\begin{equation}
M=
\begin{pmatrix}
-a & b\\
c & -c
\end{pmatrix},
\end{equation}
with
\begin{equation}
a=\frac{K_\gamma+K_\infty+K_1}{\Xi},
\quad
b=\frac{K_1}{\Xi},
\quad
c=\frac{K_1}{C}.
\end{equation}
The eigenvalues satisfy
\begin{equation}
\lambda^2+(a+c)\lambda+c(a-b)=0.
\end{equation}
Thus
\begin{equation}
\lambda_\pm=-\frac{a+c}{2}\pm\frac{1}{2}\sqrt{(a-c)^2+4bc}.
\end{equation}
The ordering is
\begin{equation}
\lambda_-<\lambda_+<0.
\end{equation}
The velocity residues are
\begin{equation}
r_+=\frac{A}{\Xi}\frac{\lambda_++c}{\lambda_+-\lambda_-},
\quad
r_-=\frac{A}{\Xi}\frac{\lambda_-+c}{\lambda_--\lambda_+}.
\end{equation}
The interlacing
\begin{equation}
\lambda_-<-c<\lambda_+<0
\end{equation}
follows from \(P(-c)=-bc<0\). Hence \(r_+>0\) and \(r_->0\) for \(A>0\).

\section{Appendix E: Threshold formulas}
\label{app:threshold}

The total adhesion threshold is
\begin{equation}
w_{\mathrm{crit}}(R)=\frac{2\kappa}{R^2}+\gamma x_f+\frac{\chi E_\infty R x_f}{2\pi}.
\end{equation}
The threshold-minimizing radius is
\begin{equation}
R_*=\left(\frac{8\pi\kappa}{\chi E_\infty x_f}\right)^{1/3}.
\end{equation}
The minimum threshold is
\begin{equation}
w_{\mathrm{crit,min}}=\gamma x_f+\frac{6\kappa}{R_*^2}.
\end{equation}
The maximum long-time modulus at fixed \(R\) and \(w_{\mathrm{total}}\) is
\begin{equation}
E_{\infty,\max}
=\frac{2\pi}{\chi x_fR}
\left(
w_{\mathrm{total}}-\frac{2\kappa}{R^2}-\gamma x_f
\right).
\end{equation}
The maximum cell modulus is
\begin{equation}
E_{c,\max}
=\frac{E_vE_{\infty,\max}}{E_v-E_{\infty,\max}}.
\end{equation}
For fixed \(w_{\mathrm{total}}\), the peak of the critical cell modulus is at
\begin{equation}
R_{E,\max}=\sqrt{\frac{6\kappa}{w_{\mathrm{total}}-\gamma x_f}}.
\end{equation}

\end{document}